\documentclass[useAMS,usenatbib]{mn2e}
\usepackage{bm}
\usepackage{amssymb}
\usepackage{comment}
\usepackage{macros}
\usepackage[usenames,dvipsnames,svgnames,table]{xcolor}
\usepackage{hyperref}
\definecolor{darkblue}{rgb}{0.0,0.0,0.7}
\hypersetup{colorlinks,breaklinks,
            linkcolor=darkblue,urlcolor=darkblue,
            anchorcolor=darkblue,citecolor=darkblue}
\usepackage{graphicx}
\usepackage{amsmath}
\usepackage[outdir=./]{epstopdf}

\def\vph{\varphi}
\def\th{\theta}

\def\b{B_{L \rm} R_{L \rm}}
\def\td{\tanh\left(\frac{r-\beta c t'}{\Delta_{\rm lab}}\right)}

\title[On the internal structure of the current sheet in the pulsar wind]{On the internal
structure of the current sheet in the pulsar wind}
\author [V. V. Prokofev, L. I. Arzamasskiy and V. S. Beskin]{V. V. Prokofev$^1$, L. I. Arzamasskiy$^{2}$\thanks{E-mails:
beskin@lpi.ru, leva@astro.princeton.edu},
V. S. Beskin$^{1,3}$\footnotemark[1]\\
$^{1}$Moscow Institute of Physics and Technology, Dolgoprudny, Institutsky per., 9,
Moscow region, 141700, Russia\\
$^{2}$Department of Astrophysical Sciences, Peyton Hall, Princeton University, Princeton, NJ 08544, USA \\
$^{3}$P.N.Lebedev Physical Institute, Leninsky prosp., 53, Moscow,
119991, Russia}

\begin{document}

\date{Accepted. Received; in original form}

\pagerange{\pageref{firstpage}--\pageref{lastpage}} \pubyear{2017}

\maketitle

\label{firstpage}

\begin{abstract}
We investigate { the internal structure of the current sheet in the pulsar
wind within} force-free and two-fluid MHD approximations. Within the 
force-free approximation we obtain { general} asymptotic solution of the
Grad-Shafranov equation for quasi-spherical pulsar wind { up to the second
order in small parameter $\varepsilon = (\Omega r/c)^{-1}$. The solution allows an arbitrary latitudinal structure of the radial magnetic field, including that obtained in the numerical simulations of oblique rotators. 
It is also shown that the shape of the current sheet does not depend on the 
latitudinal structure.} 
For the internal region of the current sheet { outside the fast magnetosonic 
surface} where the force-free approximation is not valid we use two-fluid MHD approximation.  
{ Carrying out calculations in the 
comoving reference frame we succeed in determining intrinsic electric and magnetic fields 
of a sheet. It allows us to analyze time-dependent effects which were not investigated
up to now. In particular, we estimate} the efficiency of the particle 
acceleration inside the sheet. Finally, after investigating the motion of individual 
particles in the time-dependent current sheet, we find self-consistently the width of the 
sheet and its time evolution.
\end{abstract}

\begin{keywords}
Neutron stars -- radio pulsars
\end{keywords}


\section{Introduction}
\label{sect:intro}

Particle acceleration in compact astrophysical objects is the classical
problem of modern astrophysics. Indeed, high energy radiation in the GeV
and even TeV energy band is the smoking gun of the existence of the relativistic
particles with characteristic Lorentz-factors $\gamma$ up to $10^4$--$10^5$ 
\citep{2000PhRvL..85..912L, 2007A&A...464..235A}.

Radio pulsars are thought to be the most effective accelerators in space.
Indeed, fast rotation of a neutron star (radius $R \sim 10$--$15$ km and
periods $P \sim 1$ s for ordinary pulsars) with surface magnetic field
$B_{0} \sim 10^{11}$--$10^{13}$ G (and even  $10^{13}$--$10^{15}$ G for
magnetars) inevitably results in the generation of the large enough electric
field $E \sim (\Omega R_{0}/c) B_{0}$. Here $\Omega = 2 \pi/P$ is the
neutron star angular velocity, and $R_{0}$ is the effective radius of the 
`central engine'; for radio pulsars $R_{0} \approx (\Omega R/c)^{1/2} R$
is the polar cap radius. The energetics of radio pulsars is determined by the
potential drop $V_{\rm tot} \sim e E R_{0}$.

To determine self-consistently the specific realization of the effective
particle acceleration, it is necessary to know in detail the structure of the
neutron star magnetosphere and the pulsar wind. Important analytical results
were already obtained in the first quarter of the century since radio pulsars
were discovered. In particular, the role of the process of the quantum-mechanical
particle creation was clarified~\citep{sturrock71, rs75, 1981ApJ...248.1099A}. 
It was shown the very possibility of the quasi-radial magnetized wind transporting 
electromagnetic energy to infinity~\citep{1973ApJ...180L.133M, 1975MNRAS.170..551B, 
1999A&A...349.1017B}. It was also predicted the full screening of magnetodipole 
radiation by plasma filling the neutron star magnetosphere~\citep{BGI}. Later 
these results were confirmed with numerical simulations of axisymmetric~\citep{ckf99, 
2006MNRAS.368.1055T, 2005PhRvL..94b1101G, 2006MNRAS.367...19K} as well as inclined 
magnetosphere~\citep{2012MNRAS.420.2793K, SashaMHD}.

As a result, at large enough distance from a neutron star $r \gg R_{\rm L}$
($R_{\rm L} = c/\Omega$ is the radius of the light cylinder) the theory predicts
the quasi-radial outflow of the relativistic electron-positron plasma along the
poloidal magnetic field. As the total flux of the magnetic field
through the whole sphere is to vanish, such a structure is to contain the current
sheet separating outgoing and ingoing magnetic fluxes. Up to the light cylinder the
energy is mainly transported by the electromagnetic field (i.e., by the flux of the
Poynting vector ${\bf S} = (c/4\pi){\boldsymbol E} \times {\boldsymbol B}$), but somewhere at larger
distances the electromagnetic energy flux is to be transferred into the particle energy.
Actually, the mechanism of this transformation is the main subject of the particle
acceleration theory. Up to now it remains the open question as the ideal MHD predicts
very ineffective particle acceleration for quasi-radial outflow~\citep{1994PASJ...46..123T, bes98}.

Clearly, for the MHD outflow to exist, the electric field ${\boldsymbol E}$ has to be smaller
 than the magnetic field ${\boldsymbol B}$. In the pulsar wind it can take place only if the
total electric current $I$ flowing in the magnetosphere (and producing toroidal
magnetic field $B_{\varphi}$) is large enough so that at the light cylinder the
toroidal magnetic field $B_{\varphi} \approx 2I/R_{\rm L}c$ becomes as large
as the electric field. Indeed, both $B_{\varphi}$ and $E$ diminish with
the distance as $r^{-1}$ (and, accordingly, $S \propto r^{-2}$). As the poloidal
magnetic field $B_{\rm p}$ for quasi-spherical structure decreases much faster
($B_{\rm p} \propto r^{-2}$), the very MHD approximation $E < B$ can be fulfilled
only if the total electric current  $I$ circulating in the pulsar magnetosphere is
as large as the so-called Goldreich-Julian current
\begin{equation}
I_{\rm GJ} = \pi R_{0}^2 j_{\rm GJ}^{\rm A},
\label{IGJ}
\end{equation}
where
\begin{equation}
j_{\rm GJ}^{\rm A} = \frac{\Omega B}{2 \pi}
\label{jGJA}
\end{equation}
is the amplitude of the Goldreich-Julian current density.

Here it is necessary to stress one important point. The definition of the Goldreich-Julian
charge density
\begin{equation}
\rho_{\rm GJ} = -\frac{\bf{\Omega} {\boldsymbol B}}{2 \pi c}
\label{rhoGJ}
\end{equation}
contains the factor $\cos \theta_{\rm m}$, where $\theta_{\rm m}$
is the angle between the vectors ${\bf \Omega}$ and ${\boldsymbol B}$. As a result,
for the condition (\ref{IGJ}) to be fulfilled for large enough inclination
angles $\chi$ between ${\bf \Omega}$ and the neutron star magnetic moment
${\bf m}$, the current density $j$ is to be larger than the local GJ current
density $j_{\rm GJ} = \rho_{\rm GJ} c$ near magnetic poles where 
$\theta_{\rm m} \sim \chi $, and
\begin{equation}
j_{\rm GJ} \approx \frac{3}{2} \, \frac{\Omega B}{2 \pi} \, \cos\theta.
\label{jjGJ}
\end{equation}

Thus, for quasi-radial MHD outflow to exist, the pair creation mechanism
has to support large enough longitudinal current $j > j_{\rm GJ}$ (\ref{jjGJ}).
At present, the most scientists believe in this model~\citep[see, e.g.,][]{2013MNRAS.429...20T},
and here we follow this point of view as well. On the other hand, effective
particle acceleration can be realized if the particle generation mechanism
cannot support large enough longitudinal current $j$. It can take place
in the vicinity of the so-called light surface $E = B$ which inevitable
appears outside the light cylinder for $j < j_{\rm GJ}$ ~\citep{BGI, 2000MNRAS.313..433B}.
Recently, the possible existence of such effective particle acceleration region was demonstrated by 
analyzing the TeV time-dependent radiation of the Crab pulsar~\citep{2012Natur.482..507A}.

Returning to standard model of the pulsar wind, let us recall that it contains `striped' 
current sheet separating ingoing and outgoing magnetic fluxes. This current sheet was predicted 
analytically~\citep{coroniti_striped_wind_1990, michel_pulsar_wind_1994, 1999A&A...349.1017B}
and later was confirmed in numerous force-free~\citep{ckf99, 2006ApJ...648L..51S, 2012MNRAS.420.2793K}, 
MHD~\citep{2006MNRAS.367...19K, 2009ApJ...699.1789T} and even PIC~\citep{2015MNRAS.448..606C}  
simulations\footnote{Nowhere any restrictions on the longitudinal current were imposed.}.
And it has long been understood that this current sheet can be the domain of very effective particle
acceleration~\citep{2001ApJ...547..437L,2007ApJ...670..702Z,2012SSRv..173..341A,2014ApJ...781...46C} 
On the other hand, self-consistent model of the internal structure of this 'striped' current sheet
was not constructed. 

The paper is organized as follows. We start with the discussion of force-free asymptotic 
behavior of the pulsar wind in Sect.~\ref{sect:rest}.  Here we obtain simple asymptotic 
solution of the Grad-Shafranov equation for quasi-spherical pulsar wind. { In 
particular, this solution can describe the latitudinal structure of the radial magnetic field obtained 
numerically for oblique rotator~\citep{2016MNRAS.457.3384T}}. We also show that the shape 
of the current sheet does not depend on the latitudinal structure. Then in Sect.~\ref{sect:fields} 
we determine the main properties of the internal regions of a current sheet where the force-free
approximation is not valid.  Using two-fluid MHD approximation and carrying out calculations 
in the comoving reference frame we determine electric and magnetic field structure as well as 
the velocity component perpendicular to the sheet. This allows us to estimate the efficiency 
of particle acceleration. After that we find the self-consistent solution for the current sheet evolution. 
Sect.~\ref{sect:IntStruct} is devoted to numerical simulation which, as will be shown, 
fully support our analytical asymptotic solutions. Finally, in Sect.~\ref{sect:conc} we discuss 
possible astrophysical applications of our consideration.

\section{Force-free asymptotic behavior of the pulsar wind}
\label{sect:rest}

\subsection{Basic equations}

In this section we discuss the asymptotic behavior of the pulsar wind
using well-known approach of the so-called \textit{pulsar equation}
~\citep{1973Ap&SS..24..289M, 1974MNRAS.167..457O}
\begin{equation}
\begin{split}
-\left(1-\frac{\Omega_{\rm F}^2\varpi^2}{c^2}\right)\nabla^2\Psi
+2\frac{1}{\varpi}\frac{\partial\Psi}{\partial\varpi}
+\frac{\varpi^2\Omega_{\rm F}}{c^2}(\nabla\Psi)^2\frac{\mathrm{d}\Omega_{\rm F}}{\mathrm{d}\Psi}-
\\
-\frac{16\pi^2}{c^2}I\frac{\mathrm{d}I}{\mathrm{d}\Psi} = 0.
\label{G-SH}
\end{split}
\end{equation}
This equation resulting directly from Maxwell equations
describes axisymmetric stationary electromagnetic fields
\begin{eqnarray}
\bm{E} & = & -\frac{\Omega_{\rm F}}{2\pi c}\nabla\Psi,
\label{fieldsE} \\
\bm{B} & = & \frac{\nabla\Psi\times {\bf e}_{\varphi}}{2\pi r\sin\theta}
-\frac{2I}{cr\sin\theta} {\bf e}_{\varphi}
\label{fieldsB}
\end{eqnarray}
within the force-free approximation. Here $\Psi=\Psi(r,\theta)$ is the magnetic
flux, and two integrals of motion $\Omega_{\rm F}=\Omega_{\rm F}(\Psi)$ 
and $I=I(\Psi)$ are the so-called field angular velocity (more exactly, the angular
velocity of a test charged particle drifting in the crossed electromagnetic fields) 
and the total current inside the magnetic tube respectively. Below for simplicity we  
consider the case $\Omega_{\rm F} = \Omega$ only. It corresponds to the fast enough 
rotation of the neutron star when the potential drop in the inner gap $V_{\rm gap}$ 
is much smaller than the maximum value $V_{\rm tot}$.

The first solution of the pulsar equation (\ref{G-SH}) containing radial wind
was obtained by \citet{1973ApJ...180L.133M}. He demonstrated that the `split 
monopole' magnetic field corresponding to magnetic flux
\begin{eqnarray}
\Psi(r, \theta) & = & \Psi_{\rm tot}(1 - \cos\theta), \qquad \theta < \pi/2, \\
\Psi(r, \theta) & = & \Psi_{\rm tot}(1 + \cos\theta), \qquad \theta > \pi/2
\label{Michel}
\end{eqnarray}
is the exact solution of the pulsar equation if the additional relation
\begin{equation}
4 \pi I(\Psi) = \Omega_{\rm F} \left(2\Psi - \Psi^2/\Psi_{\rm tot}\right)
\label{EeqB1}
\end{equation}
holds. In this case
\begin{eqnarray}
B_{r} & = & B_{\rm L}\frac{R_{\rm L}^2}{r^2} \, {\rm sign}(\cos\theta),
\label{m1973'} \\
B_{\varphi} & = & E_{\theta} = -B_{\rm L}\frac{R_{\rm L}}{r}\sin\theta 
\, {\rm sign}(\cos\theta),
\label{m1973}
\end{eqnarray}
where $B_{\rm L} = B_{\rm p}(R_{\rm L}, \pi/2)$. 
This solution describing the axisymmetric case is called the `split-monopole' one as it contains 
the current sheet in the equatorial plane separating ingoing and outgoing magnetic fluxes.

Later~\citet{1973ApJ...186..625I} has found more general asymptotic solution, i.e., 
the solution in the limit $r \rightarrow \infty$. He has shown that in this limit 
an arbitrary function $\Psi = \Psi(\theta)$ remains the solution of the pulsar 
equation (\ref{G-SH}) if
\begin{equation}
4 \pi I(\theta) = \Omega_{\rm F} \sin\theta \, \frac{{\rm d}\Psi}{{\rm d}\theta}.
\label{EeqB2}
\end{equation}
According to (\ref{fieldsB}), this implies that in the asymptotic limit
$r \rightarrow \infty$ any $\theta$-dependence of the poloidal magnetic field
$B_{\rm p} = B_{\rm p}(\theta)$ can be realized. Here we have to stress that 
relation (\ref{rhoGJ}) for charge density remains true for monopole structure 
(\ref{m1973'})--(\ref{m1973}) only. In general case
\begin{equation}
\rho_{\rm e} = -\frac{\Omega_{\rm F}}{4 \pi \sin\theta} \,
\frac{{\rm d}}{{\rm d}\theta}\left(\sin\theta\frac{{\rm d}\Psi}{{\rm d}\theta}\right).
\end{equation}
As one can easily check using Eqns. (\ref{fieldsE})--(\ref{fieldsB}), both conditions 
(\ref{EeqB1}) and (\ref{EeqB2}) correspond to the clear relation $E_{\theta} = B_{\varphi}$.

Finally, it was found that the appropriate solution can be constructed for the oblique
rotator as well. According to~~\citet{1999A&A...349.1017B}, the `inclined split monopole'
\begin{eqnarray}
B_{\rm p} & = & B_{\rm L}\frac{R_{\rm L}^2}{r^2} \, {\rm sign}(\Phi), \\
B_{\varphi} & = & E_{\theta} = -B_{\rm L}\frac{R_{\rm L}}{r}\sin\theta \, {\rm sign}(\Phi),
\label{b1999}
\end{eqnarray}
where now
\begin{eqnarray}
\Phi = \cos\theta \cos\chi - \sin\theta \sin\chi \cos\left[\varphi - \Omega \left(t - r/c\right)\right],
\label{Phi}
\end{eqnarray}
and $\chi$ is the inclination angle, is the exact solution of the pulsar equation as well.
In the polar regions \mbox{$\theta < \pi/2 - \chi$} and \mbox{$\theta > \pi/2 + \chi$}
this solution coincides with the time-independent Michel solution (\ref{m1973'})--(\ref{m1973}),
but in the equatorial region $\pi/2 - \chi < \theta < \pi/2 + \chi$ all the field components
change the signs at the current sheet locating at the position $\Phi = 0$.

\subsection{Asymptotic behavior}
\label{sect:Axis}

In this section we are going to generalize the solutions mentioned above to substantiate
the result of numerical simulation. Indeed, as was shown by~~\citet{SashaMHD}, for large enough
inclination angle $\chi > 30^{\circ}$ the $\varphi$-average poloidal magnetic field depends
on the angle $\theta$ as
\begin{eqnarray}
\langle B_{\rm p} \rangle_{\varphi} \, \propto \, \sin\theta.
\label{Bpp}
\end{eqnarray}
For this reason, the structure of the solution with $\theta$-dependent poloidal magnetic field
(\ref{EeqB2}) is to be considered in more detail.

As we are mainly interested in the asymptotic behavior $r \gg R_{\rm L}$, one can search
a solution of the pulsar equation (\ref{G-SH}) in the form
\begin{equation}
\Psi(r,\theta) = \sum\limits_{n=0}^{\infty}\Psi_{n}(\theta)
\left(\dfrac{R_{\rm L}}{r}\right)^{2n}\mathrm{sign}(\cos\theta).
\label{dPsi}
\end{equation}
As was already stressed, the function $\Psi_{0}(\theta)$ describing the asymptotic 
magnetic field can be arbitrary if the condition (\ref{EeqB2})
$4 \pi I(\theta) = \Omega_{\rm F} \sin\theta \, {\rm d}\Psi_{0}/{\rm d}\theta$ holds.
According to (\ref{fieldsE})--(\ref{fieldsB}), we obtain in the zero approximation
\begin{eqnarray}
&&B_r  =  B_{\rm L}\left(\frac{R_{\rm L}}{r}\right)^2 \,
\frac{F(\theta)}{\sin\theta} \, \mathrm{sign}(\cos\theta),
\label{BRr} \\
&&B_{\varphi} =  E_{\theta} = -B_{\rm L}\frac{R_s}{r}F(\theta) \mathrm{sign}(\cos\theta),
\label{Ett} \\
&&B_{\theta}  = E_{\varphi}=E_{r}=0,
\end{eqnarray}
where $F(\theta) = (4/\pi)\Psi_0'/\Psi_{\rm tot}$, $\Psi_{\rm tot}=\Psi_{0}(\pi/2)$, 
and primes indicate the $\theta$-derivatives, e.g., $\Psi_0' = {\rm d}\Psi_0/{\rm d}\theta$. 
The Michel monopole solution corresponds to $F(\theta) = \sin\theta$. As to equation for 
the first disturbance $\Psi_{1}(\theta)$, it looks like
\begin{align}
\frac{1}{\sin\theta}\frac{\mathrm{d}}{\mathrm{d}\theta}
\left(\sin\theta\frac{\mathrm{d}\Psi_1}{\mathrm{d}\theta}\right)
&- \left(\cot^2\th-3 +
 3\frac{F'}{F}\cot\theta+\frac{F''}{F}\right)\Psi_1 =\nonumber\\ 
&= \Psi_{\rm tot}\frac{1}{\sin\theta}
\frac{\mathrm{d}}{\mathrm{d}\theta}\left(\frac{F}{\sin\theta}\right).
\end{align}
In particular, for $F(\theta) = \sin\theta$ we obtain $\Psi_{1}(\theta) = 0$, i.e., 
pure radial flow. On the other hand, for $\Psi_{0}'(\theta) =\Psi_{\rm tot} \sin^2\theta$ (and, hence, 
$B_{r} \propto \sin\theta$) we obtain
\begin{eqnarray}
\frac{1}{\sin\theta}\frac{\mathrm{d}}{\mathrm{d}\theta}
\left(\sin\theta\frac{\mathrm{d}\Psi_1}{\mathrm{d}\theta}\right)
- \left(9\cot^2\th-5\right)\Psi_1
= 2\Psi_{\rm tot}\cot\theta.
\label{Psi1}
\end{eqnarray}


\begin{figure}
\centering
\includegraphics[scale=0.7]{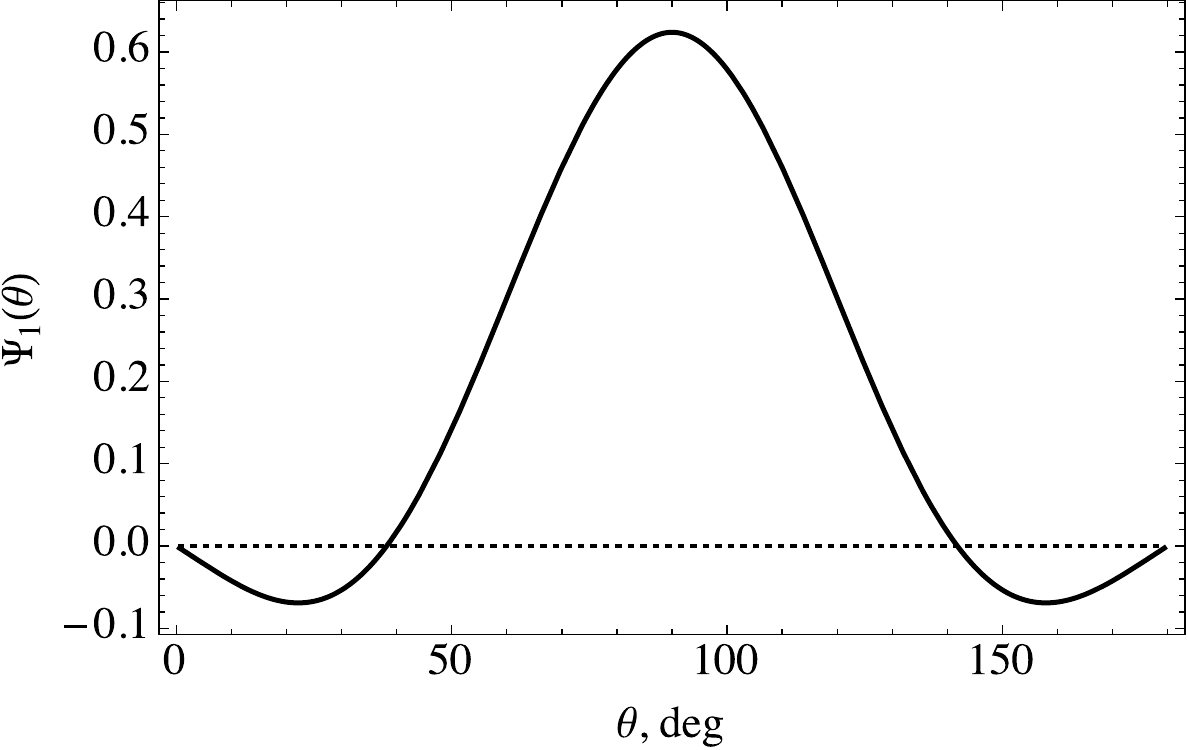}
\caption{The only solution  $\Psi_{1}(\theta)$ of equation (\ref{Psi1}) which has no singularity at $\theta = 0$ and \mbox{$\theta = \pi/2$}. The presence of finite solution implies that the disturbance of magnetic flux function decreases as $R^2_{\rm L}/r^2$.}
\label{fig:Ing}
\end{figure}

On Fig.~\ref{fig:Ing} we show the only solution $\Psi_{1}(\theta)$ of Eqn. (\ref{Psi1}) which 
has no singularity at $\theta = 0$ and $\theta = \pi/2$; these two conditions just determine 
the solution of this equation. As we see, this function is finite and, hence, the disturbance of the
radial poloidal magnetic field decreases as $R_{\rm L}^2/r^2$. We will use this property in what follows.

Now, to obtain the fields in oblique case we use the procedure similar to one
applied in~~\citet{1999A&A...349.1017B}. Instead of multiplying our axisymmetric solution by
${\rm Sign}(\cos\theta)$ we will multiply it by ${\rm sign}(\Phi)$. As one can easily check,
the appropriate fields satisfy Maxwell equations as well except for the current
sheet position $\Phi(r, \theta, \varphi) = 0$. Thus, one can conclude that the shape
of the current sheet does not depend on the latitudinal structure of the magnetic field.

Finally, using the definitions (\ref{BRr}) and (\ref{Ett}), one can obtain the
following very simple relation between the latitudinal structure of the radial
magnetic field $B_{r}(\theta)$ and the pulsar wind energy flux
$S(\theta)=cB_{\varphi}E_{\theta}/4\pi$ 
\begin{equation}
S(\theta) \propto \sin^2\theta B_{r}^2(\theta).
\end{equation}
The same expression for Pointing flux was also obtained by~\citet{2016MNRAS.457.3384T}. For $B_{r} =$ const we return to the expression $S(\theta) = \sin^2\theta$
which was widely used in the literature~\citep{2002AstL...28..373B, 2003MNRAS.344L..93K}.
On the other hand, for
$B_{r}(\theta)  \propto  \sin\theta$ we have
\begin{equation}
S(\theta) \propto \sin^4\theta,
\end{equation}
\\
i.e., exactly what was obtained numerically by~\citet{SashaMHD}. Of course, as in Eqn.
(\ref{Bpp}), it concerns $\varphi$-averaging values. Nevertheless, this result
implies that even very simple analytical consideration provide good enough description of
the main characteristics of the pulsar wind obtained numerically.

\section{Current sheet in the comoving reference frame}
\label{sect:fields}

\subsection{MHD approximation}

There are two reasons why the force-free model considered above is
too simple to describe the main properties of the current sheet. First, in
this model the sheet is infinitely thin. Second, within the force-free
approximation massless particles move with the velocity of light. As one
can see from (\ref{Phi}), the current sheet moves with the same velocity as
well, which prevents us considering in detail its internal structure.

For this reason below we try to pass to the reference frame comoving with the
outflowing plasma. It help us to separate the intrinsic processes inside the
current sheet from the common outflowing motion. Certainly, within the force-free
approximation this boost is impossible. For this reason in what follows we use
more general MHD approximation formulated in ~\citet{2000MNRAS.313..433B} 
\citep[see also][]{beszak04}.

Remember that to describe magnetically dominated MHD outflow it is very convenient
to introduce two dimensionless parameters, namely Michel magnetization parameter
$\sigma_{\rm M}$ and multiplicity parameter $\lambda$ ~\citep{2010mfca.book.....B}. Here
\begin{equation}
\sigma_{\rm M} = \frac{e\Omega \Psi_{\rm tot}}{8 \pi \lambda m_{\rm e}c^3},
\label{sigma}
\end{equation}
where $\Psi_{\rm tot} = \Psi(\pi/2)$ is the total magnetic flux through the upper
hemisphere, and
\begin{equation}
\lambda = \frac{n_{\rm e}}{n^{0}_{\rm GJ}},
\end{equation}
where $n^{0}_{\rm GJ} = \Omega B_{\rm p}/2\pi c |e|$ is the amplitude of the Goldreich-Julian 
number density. For ordinary pulsars \mbox{$\sigma_{\rm M} \sim 10^3$--$10^4$,} and 
$\lambda \sim 10^3$--$10^4$, and only for the fast young pulsars (Crab, Vela)
\mbox{$\sigma_{\rm M} \sim 10^5$--$10^6$,} and $\lambda \sim 10^4$--$10^5$. In the force-free 
limit $m_{\rm e} \rightarrow 0$ we have $\sigma_{\rm M} \rightarrow \infty$. On the other hand, 
for finite $\sigma_{\rm M}$ the particle velocity is smaller than that of light (see below). In 
addition, we suppose that the injection Lorentz-factor $\gamma_{\rm in} \sim 10^2$ is constant 
for all outflowing region.

As a result, according both to the theory~\citep{bes98, beam2015} and
numerical simulation~\citep{1999MNRAS.305..211B, 2006MNRAS.368.1717B},
the quasi-radial MHD flow is to intersect the fast magnetosonic surface at the distance
\begin{equation}
r_{\rm F} =  {\rm min}(\sigma_{\rm M}^{1/3} \sin^{-1/3}\theta, 
\sqrt{\sigma_{\rm M}/\gamma_{\rm in}}) \, R_{\rm L},
\label{rF}
\end{equation}
the Lorentz-factor at this surface being
\begin{equation}
\gamma_{\rm F} = {\rm max}(\sigma_{\rm M}^{1/3}\sin^{2/3}\theta, \gamma_{\rm in}).
\label{gF}
\end{equation}
For $\gamma_{\rm in} \gg \sigma_{\rm M}^{1/3}$ (slow rotation) it is necessary to use the
second expressions, while for  $\gamma_{\rm in} \ll \sigma_{\rm M}^{1/3}$ (fast rotation)
they realized in the narrow cone $\theta < \gamma_{\rm in}^{3/2}\sigma_{\rm M}^{-1/2}$ near
the rotational axis only; for most angles the first expressions are to be used.

It is very important that for both fast and slow rotation outside the fast magnetosonic
surface there are actually no collimation and particle acceleration. More exactly,
\begin{equation}
\gamma \approx \sigma_{\rm M}^{1/3}\sin^{2/3}\theta \log^{1/3}(r/r_{\rm F})
\end{equation}
for fast, and $\gamma_{\rm F} \approx \gamma_{\rm in}$ for slow 
rotation~\citep{1994PASJ...46..123T, bes98, beam2015}. This implies that 
with the logarithmic precision one can believe that at large 
distances $r \gg r_{\rm F}$ the particles move with constant 
velocity~\mbox{$v_{r} < c$} exactly corresponding to the drift 
velocity in~\mbox{$U_{\rm dr}/c = E/B$}~\citep{2009ApJ...699.1789T, 2010mfca.book.....B}. 
This property just helps us to move into the reference frame comoving with a particular 
part of the wind (at some constant $\theta$).

\subsection{Fields in the comoving reference frame}

\subsubsection{Introductory remarks}
\label{sect:rem}

As was already stressed, the first attempts in describing the internal structure 
of the `striped' pulsar wind were done by~\citet{coroniti_striped_wind_1990} and ~\citet{michel_pulsar_wind_1994}. They based their analysis on magnetic reconnection. 
%

The next step was made by~\citet{2001ApJ...547..437L} describing the internal structure 
of the moving current sheet by introducing `fast' and `slow' variables allowing to consider 
the current sheet moving radially with the velocity \mbox{$v_{\rm sh} < c$.} { But they 
did not consider internal structure of a sheet postulating actually zero magnetic field 
inside it.} Later, the following electromagnetic fields were considered~\citep{2013MNRAS.434.2636P}
\begin{eqnarray}
B_{r}(r, \theta, \varphi, t')   & = &   B_{\rm L}\frac{R_{\rm L}^2}{r^2} \, 
\tanh\left(\dfrac{R_{\rm L}}{\Delta_{\rm lab}}\Phi_{1}\right), \\
B_{\varphi}(r, \theta, \varphi, t')  & = &   -\frac{B_{\rm L}}{\beta}
\frac{R_{\rm L}}{r}\sin\theta \, \tanh\left(\dfrac{R_{\rm L}}{\Delta_{\rm lab}}\Phi_{1}\right), \\
E_{\theta}(r, \theta, \varphi, t')   & = & = 
-B_{\rm L}\frac{R_{\rm L}}{r}\sin\theta \, \tanh\left(\dfrac{R_{\rm L}}{\Delta_{\rm lab}}\Phi_{1}\right).
\label{b1999}
\end{eqnarray}
Here now
\begin{eqnarray}
\Phi_{1} =\cos\theta \cos\chi
- \sin\theta \sin\chi \cos\left[\varphi - \Omega \left(t' - r/\beta c\right)\right],
\label{Phi1}
\end{eqnarray}
where $t'$  is the time in the laboratory reference frame, $\beta = v_{\rm sh}/c <1$ and 
the fuction $\tanh(...)$ was taken for clear historical reason{\footnote{This function 
corresponds to well-known \citet{harris1962} solution.}} 
(this function can be arbitrary).

Unfortunately, these fields cannot be considered as good enough zeroth approximation as
they have no force-free limit inside the sheet with finite thickness. Indeed, as can
be easily checked, the $\theta$ and $\varphi$ dependencies of the function $\Phi_{1}$
(\ref{Phi1}) denies the existence of $\theta$ and $\varphi$ components of the current
density inside the sheet even in the force-free approximation. On the other hand, in
this limit $j_r = \rho_{\rm e}c$, and, hence, to support the components $j_{\theta}$
and $j_{\varphi}$ the particle velocity is to be larger than that of light. By the way,
~\citet{2001ApJ...547..437L} did not include into consideration the radial component of 
the Maxwell equation $\nabla \times {\boldsymbol B} = \dots$  (it was postulated that the radial
component of the current density $j_{r} = 0$ in spite of $[\nabla \times {\boldsymbol B}]_{r} \neq 0$),
so their analysis cannot be considered as self-consistent as well.

Here we present another approach to this problem { carrying out calculations 
in the reference frame moving radially with the current sheet}. It allows us to avoid the
leading components of electromagnetic fields resulting in the common drift motion.
Our consideration is based on the another exact force-free solution obtained 
by~\citet{2011PhRvD..83l4035L}
\begin{eqnarray}
&&B_r= B_{\rm L} \left(\frac{R_{\rm L}}{r} \right)^2;
\label{eqn2_1}\\
&&B_\vph= E_\th =-\frac{\b}{r} \sin \th f(r- c t'),
\label{eqn2_2}
\end{eqnarray}
where $f(...)$ is again an arbitrary function. Having no dependence on $\theta$ and $\varphi$,
this fields are in agreement with the condition ${\boldsymbol j} = \rho_{\rm e}c \, {\bf e}_{r}$. We can
use this solution for inclined rotator because the shape of the current sheet in this case is
similar to the spherical wave, see, e.g.,~\citet{2012MNRAS.420.2793K}.

Of course, it is necessary to stress that this solution contains no sign change of the radial
component of the magnetic field $B_r$. On the other hand, as was already mentioned above, 
outside the fast magnetosonic surface $r \gg r_{\rm F} \sim \sigma_{\rm M}^{1/3} R_{\rm L}$ 
(\ref{rF}) both the disturbance of the monopole poloidal magnetic field resulting from the 
MHD disturbances~\citep{bes98} and the disturbances (\ref{dPsi}) connecting with 
$\theta$-dependence of the poloidal field have the same smallness $\sim \sigma_{\rm M}^{-2/3}$ 
at the fast surface and, hence, can be neglected.

For this reason, in what follows we put $B_r = 0$. As is well-known, this approximation is 
good enough outside the fast magnetosonic surface $r > r_{\rm F}$ and widely used in analysis 
of the pulsar wind~\citep{2001ApJ...547..437L, 2014PhRvD..89j3013B}. Indeed, according to 
(\ref{rF})--(\ref{gF}), for fast rotator in the comoving reference frame the toroidal 
magnetic field $B'_{\varphi} = B_{\varphi}/\gamma$ becomes larger than poloidal one 
$B'_{r} = B_{r}$ just outside the fast surface; for slow rotator it takes place even 
at smaller distances. 

As a result, we can modify now this solution for arbitrary
$\theta$-dependence of the poloidal magnetic field $B_{\rm r}(\theta)$ which, as was 
already stressed, better corresponds to the real structure of the pulsar wind.
For $f\equiv - \tanh$ the fields in the laboratory frame 
($r, \theta, \varphi, t^{\prime}$) can be presented as
\begin{eqnarray}
&&B_\vph (r, \theta, t') = \frac{1}{\beta} \frac{\b}{r} F(\theta) \td,
\label{F1}\\
&&E_\th (r, \theta, t') = \frac{\b}{r} F(\theta) \td\label{F2},
\end{eqnarray}
where $\Delta_{\rm lab}$ is a current sheet thickness in the laboratory reference frame, 
and $F(\theta)$ now is the arbitrary function. As was already stressed, the parameter 
$\beta = E_{\theta}/B_{\varphi}$ can be considered here as a constant.


Accordingly, the charge density in this frame is equal to Goldreich-Julian charge density:
\begin{equation}
\rho_{\rm e} = 
- \frac{\b}{4\pi r^2 \sin \theta} \frac{{\rm d}[F(\theta)\sin\theta]}{{\rm d}\theta} \td 
=  n_{GJ \rm} e,
\end{equation}
but the current density is now equal to
\begin{equation}
j = \frac{\rho_{\rm e}}{\beta}.
\end{equation}
This implies that within the MHD approximation the velocities of electron and positron
components are to be different. For magnetically dominated case one can seek the first 
order corrections in the following form:
\begin{equation}
v_r^\pm/c = 1 - \xi_r^\pm;~v_\th^\pm/c=\xi_\th^\pm;~v_\vph^\pm /c= \xi_\vph^\pm.
\end{equation}
As the particle number densities can be now written as
\begin{equation}
n^\pm = n_{GJ\rm} \left[\lambda \mp \frac{1}{4} {\hat D}_{\theta}F(\theta) \right],
\end{equation}
where ${\hat D}_{\theta}F(\theta)=F'(\theta) + F(\theta)\cot\theta$ 
(e.g., ${\hat D}_{\theta} \sin\theta = 2 \cos\theta$),
one can obtain using the definition of the current density 
\mbox{$j = e n^+ v_r^+ - e n^- v_r^-$}
\begin{equation}
\xi_r^\pm = 1-\beta \pm \frac{1}{2 \lambda \gamma^2 \beta},
\end{equation}
or
\begin{equation}
v_r^\pm/c = \beta \mp \frac{1}{2 \lambda \gamma^2 \beta}.
\end{equation}
Here $\gamma = (1-v^2/c^2)^{-1/2}$ is the Lorentz factor and $\lambda$ again is the
multiplicity parameter. 

\subsubsection{Comoving reference frame --- orthogonal case}
\label{sect:cmframe}

\begin{figure}
\centering
\includegraphics[scale=0.45]{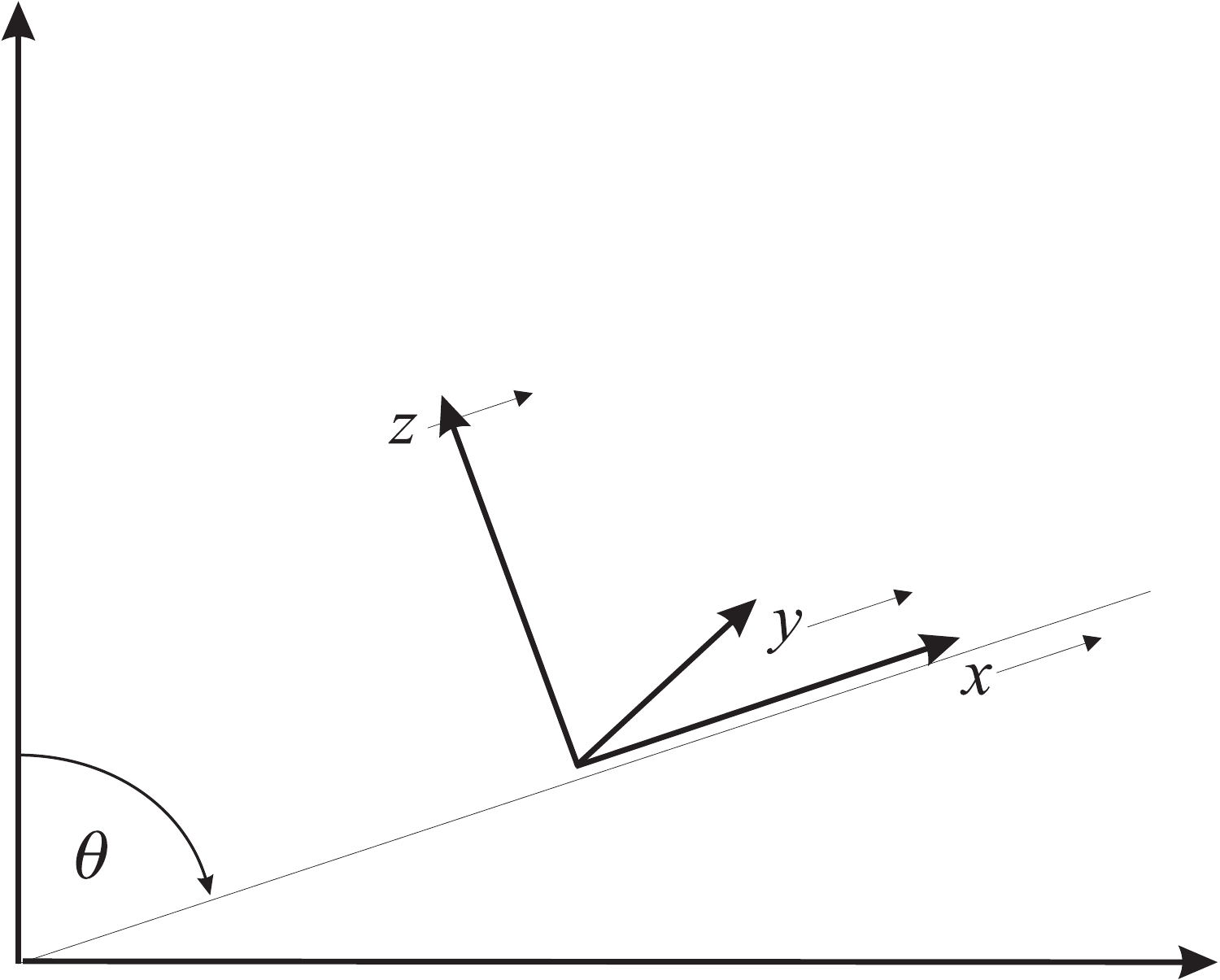}
\caption{Inertial reference frame ($x, y, z, t$) moving radially with velocity $V =\beta c$.}
\label{Fig02}
\end{figure}

As was already stressed, important property of the solution (\ref{F1})--(\ref{F2}) 
presented above is that the parameter $\beta$ can be considered as a constant. Thus, the
reference frame moving radially with the velocity $V = \beta c$ is the inertial one. 
In order to study the field structure in this reference frame moving with the current sheet,
we need to express the fields in Cartesian coordinates and make a Lorentz transform:
\begin{eqnarray}
B_x'=B_x;~B_y'=\Gamma (B_y+ \beta E_z);~B_z' = \Gamma (B_z - \beta E_y), \\
E_x'=E_x;~E_y'=\Gamma (E_y - \beta B_z);~E_z'=\Gamma(E_z+\beta B_y).
\end{eqnarray}
Here and below $\Gamma = (1-\beta^2)^{-1/2}$ is the boost Lorentz-factor, and the values 
without prime correspond to the reference frame moving radially with the $x$-axis  
directed along the radius-vector ${\bf e}_{r}$ and the $y$-axis along ${\bf e}_{\varphi}$ 
(see Fig.~\ref{Fig02}). As a result, near the origin of the reference frame moving radially with 
the velocity $V = \beta c$ (and for $f \equiv -\tanh$) the first order fields look like 
\begin{eqnarray}
B_{y}(x, y, z, t) & = & B_{0}\frac{R_{\rm L}}{c t}
\left(1 - \frac{z^2}{c^2\,t^2}\right)
\tanh \left(\frac{x}{\Gamma \Delta_{\rm lab}} \right),
\label{BYinitial} \\
E_{x}(x, y, z, t) & = & B_{0}\frac{R_{\rm L} z}{c^2t^2} \tanh
\left(\frac{x}{\Gamma \Delta_{\rm lab}} \right).
\label{EXinitial}
\end{eqnarray}
Here $x = \Gamma(r-\beta c t')$, $t = t'/\Gamma$ is the time in the comoving 
reference frame, and
\begin{equation}
B_{0} = \frac{B_{\rm L}}{\beta^2 \Gamma^2} F(\theta),
\label{B00}
\end{equation}
which now can be considered as a constant. Note that this expression 
has $\Gamma^2$ instead of $\Gamma$ in denominator since $B_0$ actually
represents magnetic field at fast magnetosonic surface in comoving 
frame. Fast magnetosonic surface corresponds to the time 
$t_0' = R_{\rm F}/\beta c \Rightarrow t_0 \approx R_{\rm L}/c$.
{ Finally, as $z \ll ct$, we do not include below the factor
$(1 - z^2/c^{2}t^{2})$ into consideration.}

Further, as one can directly check, Maxwell equation 
$c \nabla \times {\boldsymbol E} = - \partial {\boldsymbol B}/\partial t$ is automatically
fulfilled. Finally, for another Maxwell equation 
$c \nabla \times {\boldsymbol B} =  \partial {\boldsymbol E}/\partial t + 4 \pi {\boldsymbol j}$ to 
be valid, two-fluid MHD consideration is necessary. This question will be
considered in detail in Sect.~\ref{sect:IntStruct}. At the moment we only 
stress that the charge density along the boost axis (i.e., $x$-axis) is equal 
to zero, and the particle velocities after the Lorentz transform become
\begin{equation}
\frac{v_x^{\pm}}{c} = \mp \frac{1}{(2\lambda\pm 1) \beta}\approx \mp \frac{1}{2\lambda \beta}.
\end{equation}

Expressions (\ref{BYinitial})--(\ref{EXinitial}) are our main intermediate result 
giving zero approximation for the internal structure of the quasi-spherical 
current sheet in its comoving reference frame. As we see, this solution is
essentially time-dependent. Indeed, $r^{-1}$ diminishing of the toroidal 
magnetic field $B_{\varphi}$ in the pulsar wind transforms into $t^{-1}$ time
dependence in the comoving reference frame. As will be shown below, this results 
in a set of new effects which do not exist in classical time-independent
configurations.


\subsubsection{Comoving reference frame --- aligned case}
\label{sect:align}

Similarly, on can consider the internal structure of the current sheet for the
axisymmetric case when the current sheet locates in the equatorial plane. 
After passing into comoving reference frame (the calculations are quite similar 
to Section~\ref{sect:cmframe}) we obtain for the leading components of 
electromagnetic  field ($\theta = \pi/2$):
\begin{eqnarray}
B_{y}(x, y, z, t) & = & B_{0} \frac{R_{\rm L}}{ct} 
\tanh \left(\frac{z}{\Gamma \Delta_{\rm lab}} \right), \\
E_{x}(x, y, z, t) & = & B_{0} \frac{R_{\rm L}z}{c^2 t^2} 
\tanh \left(\frac{z}{\Gamma \Delta_{\rm lab}} \right).
\end{eqnarray}
This fields differ from Eqns. (\ref{BYinitial})--(\ref{EXinitial}) by $z$
(not $x$) coordinate perpendicular to the sheet plane. 

It is necessary to stress that recent PIC simulations of the axisymmetric
pulsar magnetosphere~\citep{2015MNRAS.448..606C} demonstrate non-stationarity of the equatorial
sheet so that a high amplitude wave is generated just outside the light cylinder. 
In other words, axisymmetric equatorial sheet actually cannot exist. Nevertheless,
we consider this case as well.

\section{Internal structure of time-dependent current sheet}
\label{sect:IntStruct}

In this section we discuss in detail the structure of electromagnetic fields 
and particle drift motion for time-dependent current sheet { not far 
from the fast magnetosonic surface. In this domain the time-dependent effects 
are most pronounced. Most of the section is devoted to discussion of orthogonal 
case, while aligned case is discussed in Appendix~\ref{Appendix.align}.}

{ In our analysis, we do not consider the important effects related to different instabilities (e.g., tearing and drift-kink) which can drastically disturb the structure of a sheet. The inclusion of these effects is beyond the scope of this study and we leave it for the following paper.}

Note, that for simplicity we use $f(x) = \tanh(x)$ form of the current sheet 
(Harris sheet),  everywhere except for Sect. \ref{sect.A.E.F.}. The results 
can be easily applied to any physically reasonable form, i.e. odd function with 
\begin{equation}
\lim\limits_{x\rightarrow \pm \infty}f(x)\rightarrow \pm 1 
\nonumber
\end{equation}
and  $f(0)=0$. 

\subsection{Accelerating electric field}
\label{sect.A.E.F.}

As was already stressed, the solution (\ref{BYinitial})--(\ref{EXinitial}) 
constructed above corresponds to the constant width $\Delta_{\rm lab}$ of the current 
sheet separating time-dependent magnetic fluxes. To describe the time evolution 
of the  current sheet width $\Delta_{\rm lab}(t)$ (which is one of the main goals of this paper),
it is necessary to include into accelerating another component of the electric
field $E_{z}$.

To show this, it is convenient to rewrite the relations (\ref{BYinitial})--(\ref{EXinitial}) 
in the form
\begin{eqnarray}
B_{y}(x, y, z, t) & = & B_{0}\frac{R_{\rm L}}{c t} h(x,t),
\label{BYin} \\
E_{x}(x, y, z, t) & = & B_{0}\frac{R_{\rm L} z}{c^2 t^2} h(x,t).
\label{EXin}
\end{eqnarray}
where $h(x,t)=f[x/\Delta(t)]$. Then, we can rewrite Maxwell equation 
$c \nabla \times {\boldsymbol E} = - \partial {\boldsymbol B}/\partial t$  as
\begin{equation}
\frac{\partial E_{z}}{\partial x} 
= B_{0}\frac{R_{\rm L}}{c^2 t} \frac{\partial h}{\partial t}. 
\end{equation}
It gives
\begin{equation}
E_{z}
= B_{0}\frac{R_{\rm L}}{c^2 t} \frac{\partial}{\partial t}\left(\int\limits^{x}_{\infty} h(x',t){\rm d}x'\right). 
\end{equation}
Further, as $f(x)$ and $h(x,t)$ are both odd functions of $x$, 
one can choose the integrating constant for $E_{z} \propto \int h(x',t){\rm d}x'$
to be even function with clear boundary conditions $E_{z}(\pm \infty) = 0$. Clearly, in this case 
we obtain $E_{z}(x=0)\neq 0$ near the center plane of the current sheet. Finally, Maxwell equation 
$c \nabla \times {\boldsymbol B} =  \partial {\boldsymbol E}/\partial t + 4 \pi {\boldsymbol j}$ 
now looks like
\begin{equation}
\label{eqn:ez}
B_{0}\frac{R_{\rm L}}{c t} \frac{\partial h}{\partial x} 
= \frac{1}{c \, }\frac{\partial E_{z}}{\partial t} + \frac{4 \pi}{c} \, j_{\rm z}. 
\end{equation}
For Harris current sheet one can obtain the following expressions:
\begin{align}
B_{y} & =  B_{0} \frac{t_0}{t}\tanh\dfrac{x}{\Delta(t)},
\label{ByOrt}\\
E_{x}  & =   B_{0} \frac{t_0z}{ct^2}\tanh\dfrac{x}{\Delta(t)},
\label{ExOrt}\\
E_{z} & =  B_{0} \frac{t_0\Delta'(t)}{ct}\left\{\log\left[2\cosh\dfrac{x}{\Delta(t)}\right]-
\right. \label{EzOrt} \\
&- \left.\dfrac{x}{\Delta(t)}\tanh\dfrac{x}{\Delta(t)}\right\}.
\notag
\end{align}

As we see, time-dependence of the current sheet thickness $\Delta(t)$ inevitably results in 
the appearance of the nonzero electric field $E_{z}$ along the sheet which is larger 
than magnetic one near the zero surface. Thus, the evolution of the current sheet
width and the problem of particle acceleration cannot be considered separately.

Acceleration of particles could be estimated from considering particles trapped deep inside 
the current sheet $x\ll \Delta$. For such particles, the change of $z$-component of momentum 
due to $E_z$ could be found from the solution of 
\begin{equation}
\dot p_z \approx e B_0 \frac{t_0 \Delta'(t)}{ct} \log(2),
\end{equation}
giving the acceleration
\begin{equation}
\delta p_z(t) = \int\limits_{t_0}^{t} \dot p_z {\rm d}t = \frac{\kappa \log(2)}{\kappa - 1} \frac{e B_0 \Delta_0}{c}[(t/t_0)^{\kappa -1} - 1].  
\label{eq:deltapz}
\end{equation}
Here $\Delta'(t) = {\rm d}\Delta/{\rm d} t$, and we assume power-law dependence 
for the current sheet thickness 
$\Delta(t) \propto t^\kappa$. Equation (\ref{eq:deltapz}) is valid for 
$\kappa \ne 1$. For $\kappa = 1$ integration (\ref{eq:deltapz}) gives 
logarithmic divergence. For such case we assume 
$\delta p_z \approx e B_0 \Delta_0/c = {\rm const}$. 
For $\kappa < 1$, $\delta p_z$ asymptotically goes to constant. In this 
case we also set $\delta p_z \approx e B_0 \Delta_0/c = {\rm const}$. 
On the other hand, for $\kappa > 1$ momentum grows as $t^{\kappa - 1}$
with time. Combining all the expressions, one can approximate the 
acceleration of particles in the current sheet as
\begin{align}
\delta p_z(t)  &= e B_0 \Delta_0/c, ~~~~\qquad \kappa \le 1 \label{eq:pz1}\\
\delta p_z(t)  &= e B(t) \Delta(t)/c, \qquad \kappa > 1 \label{eq:pz2}
\end{align}
where $B(t)$ is the value of magnetic field outside of the current sheet $B(t) = B_0 (t_0/t)$.

\begin{figure}
\centering
\includegraphics[scale=0.55]{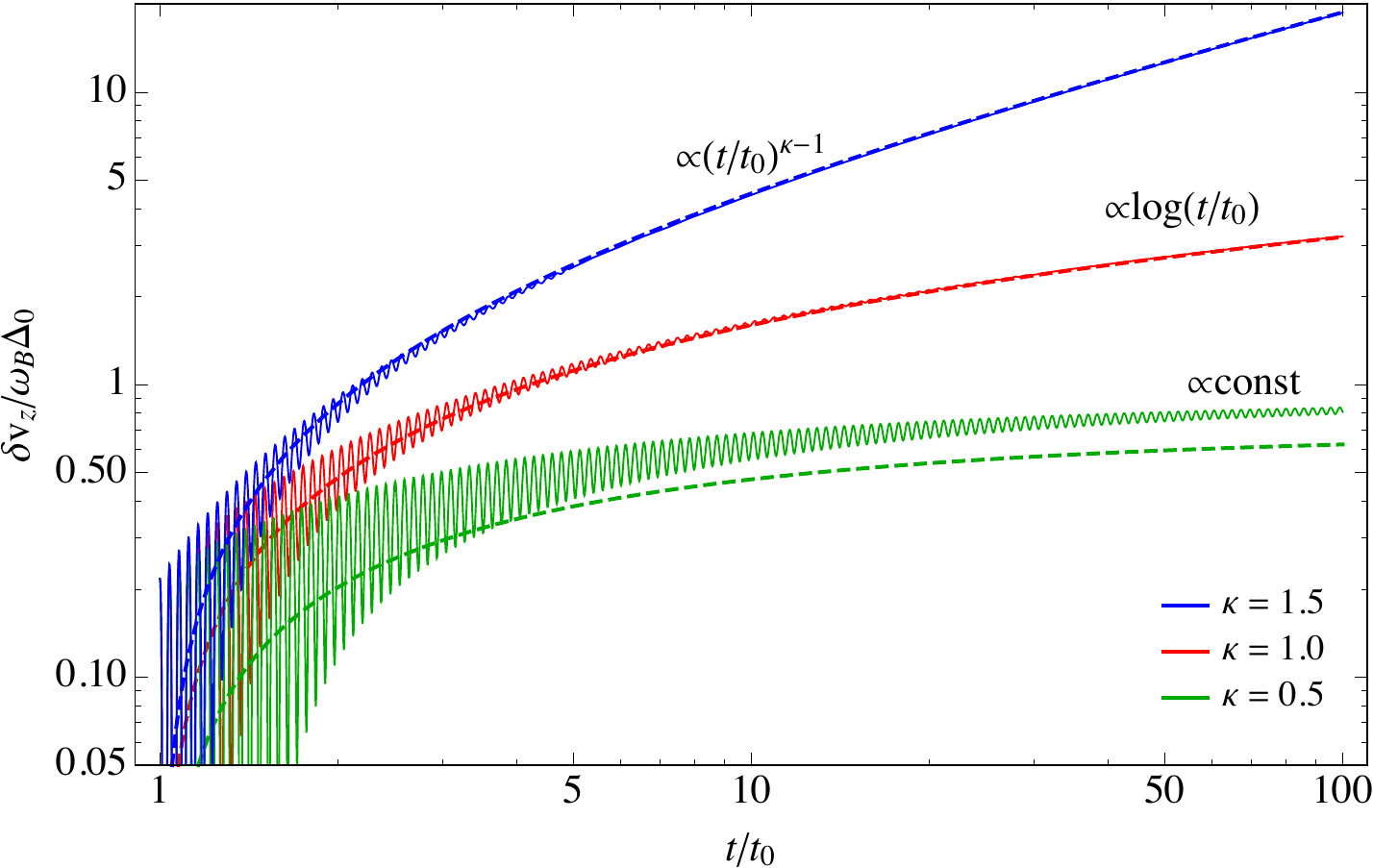}
\caption{The acceleration of particles along the sheet by an additional electric 
field for different expansion parameters $\kappa$. Solid lines correspond to the 
time dependence of particle velocity, and dashed lines represent analytical prediction 
(\ref{eq:deltapz}).}
\label{fig:p(t)}
\end{figure}

\begin{figure}
\centering
\includegraphics[scale=0.55]{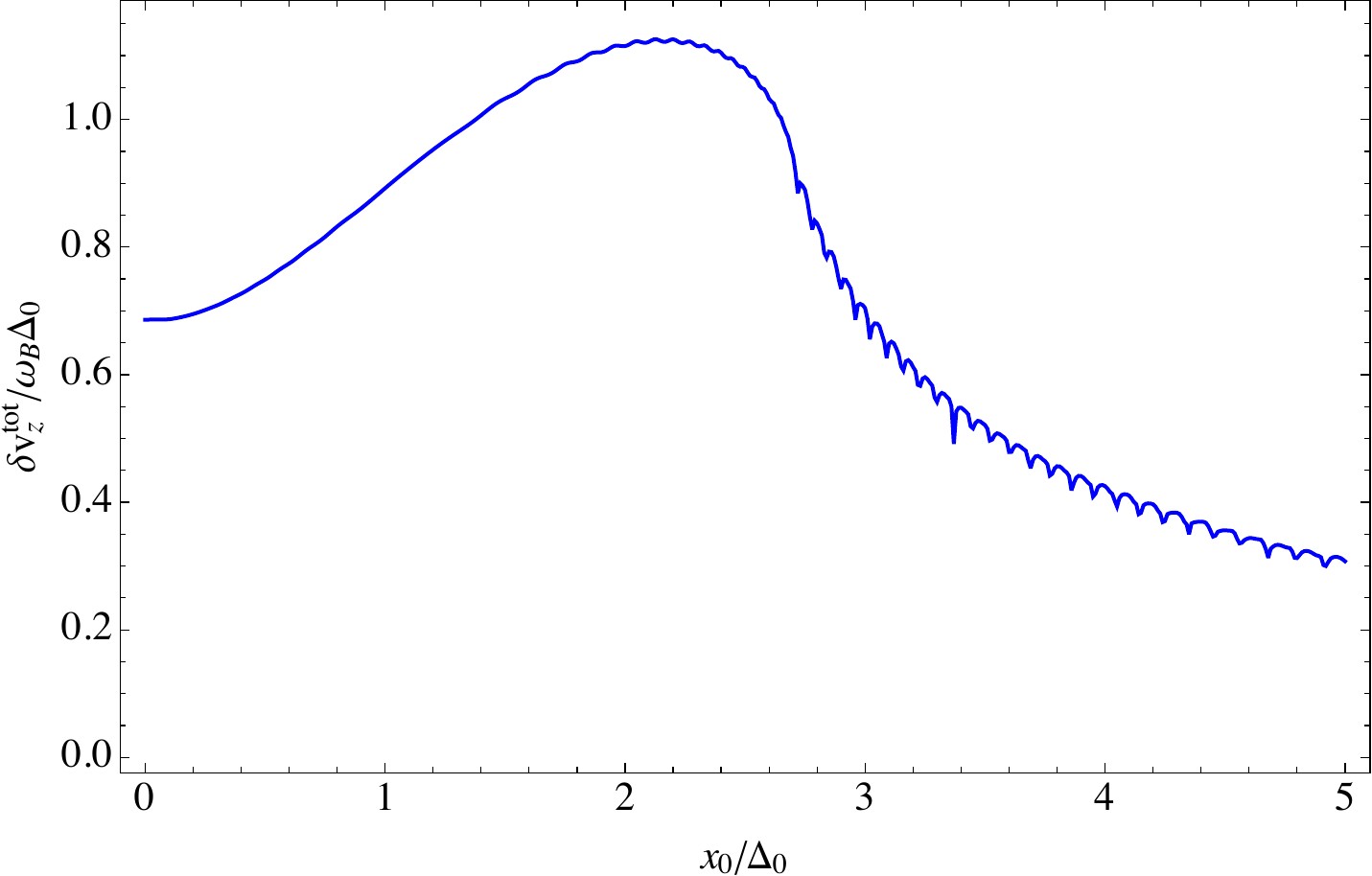}
\caption{Dependence of the total acceleration in the sheet with $\kappa = 0.5$ on the 
initial coordinate of the particle. Acceleration increases until $x_0/\Delta_0\sim 2.5$. 
It corresponds to the last initially trapped orbit. Particles with larger $x_0$ move 
outside the sheet until getting trapped, and the total change in velocity decreases.}
\label{fig:p(x)}
\end{figure}

In Figure \ref{fig:p(t)}, expression (\ref{eq:deltapz}) is compared with the exact 
solution of particle equations of motion in the fields (\ref{ByOrt})--(\ref{EzOrt})
for different $\kappa$. The numerical solution is in good agreement with analytical 
prediction, except for $\kappa = 0.5$. The small difference between analytical and 
numerical solution in $\kappa = 0.5$ case is examined in Figure \ref{fig:p(x)}. 
All particles examined for this Figure start with $ z_0 = 0$, $v_x = 0$, 
$v_z = \omega_{\rm B}\Delta_0$ and with different initial $x_0$. An analytical value 
$\delta v_z = \log 2 \omega_{\rm B}\Delta_0$ corresponds to the limit of this curve 
at $x_0 \rightarrow 0$. For larger $x_0$, acceleration is larger, but stays order of
magnitude the same. We thus conclude that estimations (\ref{eq:pz1})--(\ref{eq:pz2})
are in good agreement with numerical solution.


\subsection{Particle drift outside the current sheet}
  \label{part.drift}

{ In this section we discuss the main manifestations of time-dependent
effects in the proper reference frame of the current sheet. First of all,} in the 
comoving reference frame the particles experience drift motion outside the current 
sheet ($x \gg \Delta$ for orthogonal case and $z \gg \Delta$ for aligned case). The 
fastest component of such motion is $z$-component of electric drift 
\begin{equation}
\label{eqn.drift.vel.}
    U_z = c\frac{E_x B_y}{B_y^2} = \frac{z}{t}.
\end{equation}
This velocity exactly corresponds to the radial divergence of the flow due to 
radial expansion. Since within our approximation $z \ll ct $, the drift velocity 
of particles remains non-relativistic.

The drift velocity $U_z$ is directed along the current sheet in orthogonal case, 
and out of the sheet in aligned case. For orthogonal case it implies that the 
particle, which is outside of the current sheet initially, will move along the 
sheet until it's orbit starts to intersect the midplane of the sheet. After 
this moment the particle is trapped and no longer drifts. On the other hand, 
for aligned case, particles drift away from the sheet. If current sheet expands 
slowly enough, such particles will never get trapped. Similarly, the particles, 
which are initially trapped might escape the sheet and start to drift with 
velocity $U_{\rm esc}^{\rm align} = \Delta(t_{\rm esc})/t_{\rm esc}$, where 
$t_{\rm esc}$ is the time when particle escapes the sheet. One can easily find 
that escape can only occurred if $\Delta(t)$ expands slower than linear.

As was shown in Sect.~\ref{sect.A.E.F.}, in orthogonal case the additional time 
dependence of a sheet width leads to additional electric field diminishing 
exponentially outside the sheet: $E_{\rm add}\propto\exp(-x/\Delta)$. Since for 
particles outside the sheet $\Delta\ll x$, we can neglect this electric field. 
The same remains true for gradient drift.
 
Further, Larmor radius of particles changes due to the conservation of the first 
adiabatic invariant $p_{\perp}^2/B_y \approx$ const. This connection allows 
us to evaluate the dependence of gyroradius on time (cf.~\citealt{2001ApJ...547..437L})
\begin{equation}
\label{eqn.L.R.}
\mathcal{R}_{\rm L}(t) = \mathcal{R}_{\rm L}(t_0)\left(\dfrac{t}{t_0}\right)^{1/2}.
\end{equation}
Such a dependence shows that the gradient drift (or any other drifts except electric) 
is not just significantly smaller than $z$-component of electric drift, but 
also smaller compared to Larmor radius growth. For orthogonal case 
Eqns. (\ref{eqn.drift.vel.}) and (\ref{eqn.L.R.}) indicate that any particle that 
initially is outside the sheet eventually gets trapped inside the sheet.

Now we can compare an average gyro-radius of particles with the thickness of sheet. 
If gyroradius of a particle is much smaller than the width 
of the current sheet, it is possible to use the drift approximation. The only component 
of the drift velocity having $x$-component is the electric drift 
\mbox{$U_{x} = -c E_{z}B_{y}/B^2$} 
\begin{equation}
U_{x} = -\Delta'(t)
\left\{\dfrac{\log\left[2\cosh( x/\Delta)\right]}{\tanh (x/\Delta)}  - x/\Delta\right\},
\label{eq:Ux}
\end{equation}
 In particular, for \mbox{$x \ll \Delta_0$} we have
\begin{equation}
U_{x} = -\log(2)\frac{\Delta'(t)\Delta(t)}{x}.
\end{equation}
Certainly, it is impossible to use this evaluation in the very center of a sheet $x \rightarrow 0$
where $E_{z} > B_{y}$.

As one can see, for $\Delta \gg {\cal R}_{\rm L}$, when there are a lot of particles 
with $\mathcal{R}_{\rm L} \ll |x| < \Delta$ which do not cross the null surface, these 
particles will drift towards $x = 0$. This results in slow collapse of the sheet until 
the current sheet becomes thin enough so $\mathcal{R}_{\rm L} \sim \Delta$.

On the other hand, it is impossible to sustain the Harris current sheet if 
$\mathcal{R}_L \gg \Delta$ as the particles in such a sheet spend most of 
their time in the region $|x| > \Delta$, i.e. outside of the sheet. This 
leads to the conclusion that in realistic current sheet
$\mathcal{R}_{\rm L} \sim \Delta$.

\begin{figure}
\centering
\includegraphics[scale=1]{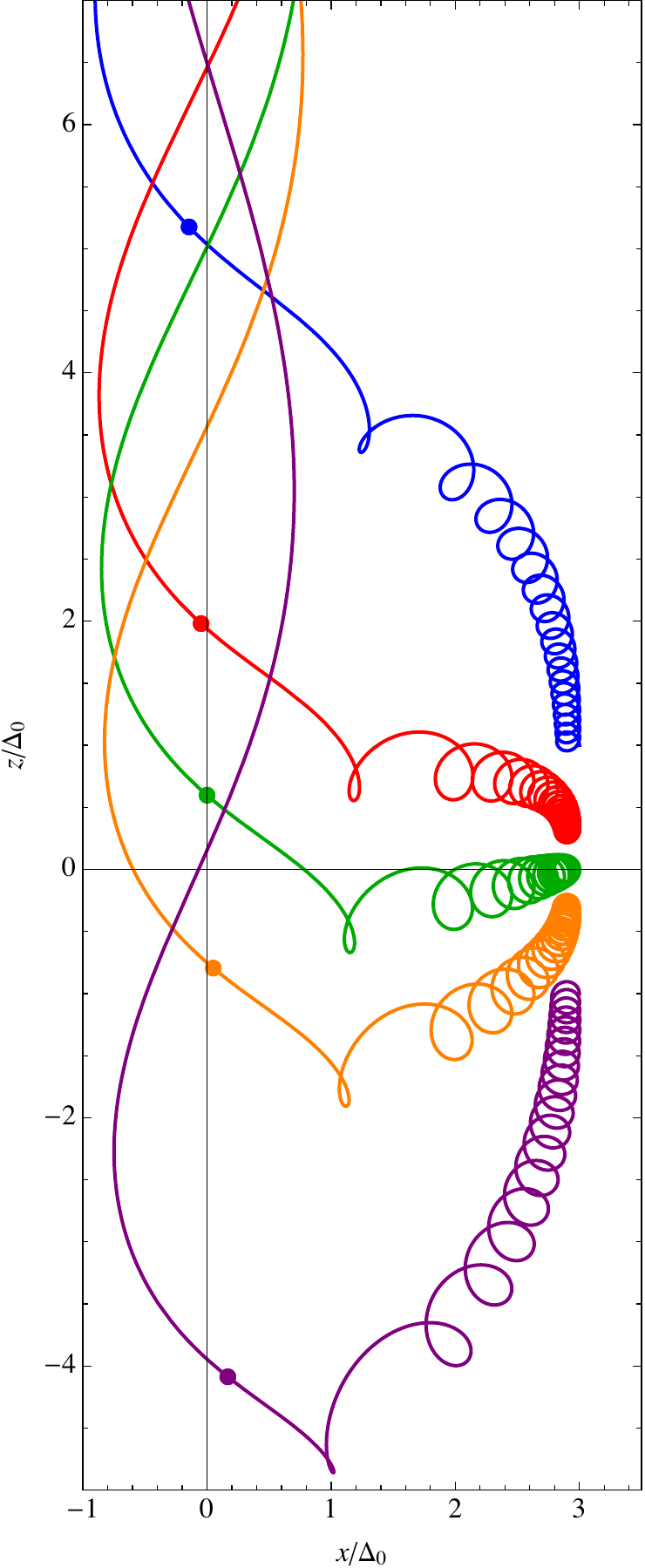}
\caption{Trapping of particles in the orthogonal current sheet for different starting 
positions outside the sheet. The points correspond to the positions of particles at 
the same moment in time. As $x$-positions of these points are close, the time until
particles get trapped is the same for particles with the same initial $x$.}
\label{fig:trap}
\end{figure}

{ Finally, time-dependence of the current sheet results in the trapping
of particles in the vicinity of the null surface. We illustrate this effect in
Figure~\ref{fig:trap} where we show} the trajectories of drifting particles outside the 
sheet. All particles start with the same velocity and $x_0$, but with different $z_0$. 
Points correspond to the positions of particles at the same moment in time showing that 
the time required for particle to get trapped is almost independent on $z_0$, and is 
determined by the $x$-component of drift velocity (\ref{eq:Ux}).

\subsection{Self-consistent solutions for non-relativistic particles}
\label{sect:solAn}

In this section we find exact solution for a particle trapped deep inside the orthogonal
current sheet $x \ll \Delta$. Surprisingly, the equations for non-relativistic particles 
could be integrated exactly. For relativistic particles not only equations could not be 
solved analytically, but also the actual approximation $x,z \ll ct$ breaks down and the 
fields (\ref{ByOrt})-(\ref{EzOrt}) could not be used.

In the orthogonal case we can use expressions (\ref{ByOrt})-(\ref{EzOrt}) for electromagnetic 
fields:
\begin{align}
B_y &= B_0 \frac{t_0}{t} \tanh\frac{x}{\Delta(t)},\\
E_x &= B_0\dfrac{t_0z}{ct^2} \tanh\frac{x}{\Delta(t)}\\
E_z &= \kappa \dfrac{B_0t_0\Delta(t)}{ct^2} 
\left\{\log\left[2\cosh\dfrac{x}{\Delta(t)}\right]-\right.\\
&-\left. \dfrac{x}{\Delta(t)} \tanh\dfrac{x}{\Delta(t)}\right\},
\end{align}
where we again assume $\Delta(t)\propto t^{\kappa}$. After substitution these expression into equation 
of motion we obtain the following system of equations:
\begin{align}
\ddot{z} &= \dfrac{eB_0t_0}{m_{\rm e} c t}\left\{\dot{x}\tanh\dfrac{x}{\Delta(t)}+\right.\\
&+\kappa\dfrac{\Delta(t)}{t}\log\left[2\cosh\dfrac{x}{\Delta(t)}\right]
-\left.\kappa\dfrac{x}{t}\tanh\dfrac{x}{\Delta(t)}\right\},
\nonumber \\
\ddot{x} &= \dfrac{eB_0t_0}{m_{\rm e} ct^2}\tanh\dfrac{x}{\Delta(t)}(z-\dot{z}t).\label{eq:ddotx}
\end{align}

Integrating now the first equation one can obtain
\begin{equation}
z-\dot{z}t=-\Delta(t)\Lambda\log\left[2\cosh\dfrac{x}{\Delta(t)}\right]+C
\label{zaz}
\end{equation}
corresponding to conversation of the following invariant 
\begin{equation}
I = m_{\rm e}z - P_{z}t.
\label{Iinv}
\end{equation}
Here $\Lambda =  \omega_{B} t_0$ ($\omega_{B}= e B_0/m_{\rm e} c$ is cyclotron frequency),
and $P_z = p_z + e A_z$ is generalized momentum. This integral remains constant for non-relativistic 
case only. Further, substituting (\ref{zaz}) into (\ref{eq:ddotx}) we obtain
\begin{equation}
\label{eq.x.full.ort}
\ddot{x}=-\Lambda^2\frac{\Delta(t)}{t^2}\tanh\dfrac{x}{\Delta(t)}\log\left[2\cosh\dfrac{x}{\Delta(t)}\right].
\end{equation}
Here we neglect the constant $C$ in comparison with $\Delta(t)$ which is possible for expanding sheet.

To evaluate the asymptotic behaviour of nonlinear equation (\ref{eq.x.full.ort}) (which has 
no exact analytical solution) we present the coordinate $x(t)$ in the form $x(t) = \Delta(t)S(t)$ 
where $S(t)$ is a restricted function. After substituting this into equation (\ref{eq.x.full.ort}) 
we obtain
\begin{equation}
t^{2} S''+2 \kappa t S' + \kappa(\kappa-1)S
= -\Lambda^2\tanh S\log[2\cosh S],
\end{equation} 
where primes indicate derivatives over $t$. 

\begin{figure}
\centering
\includegraphics[scale=0.55]{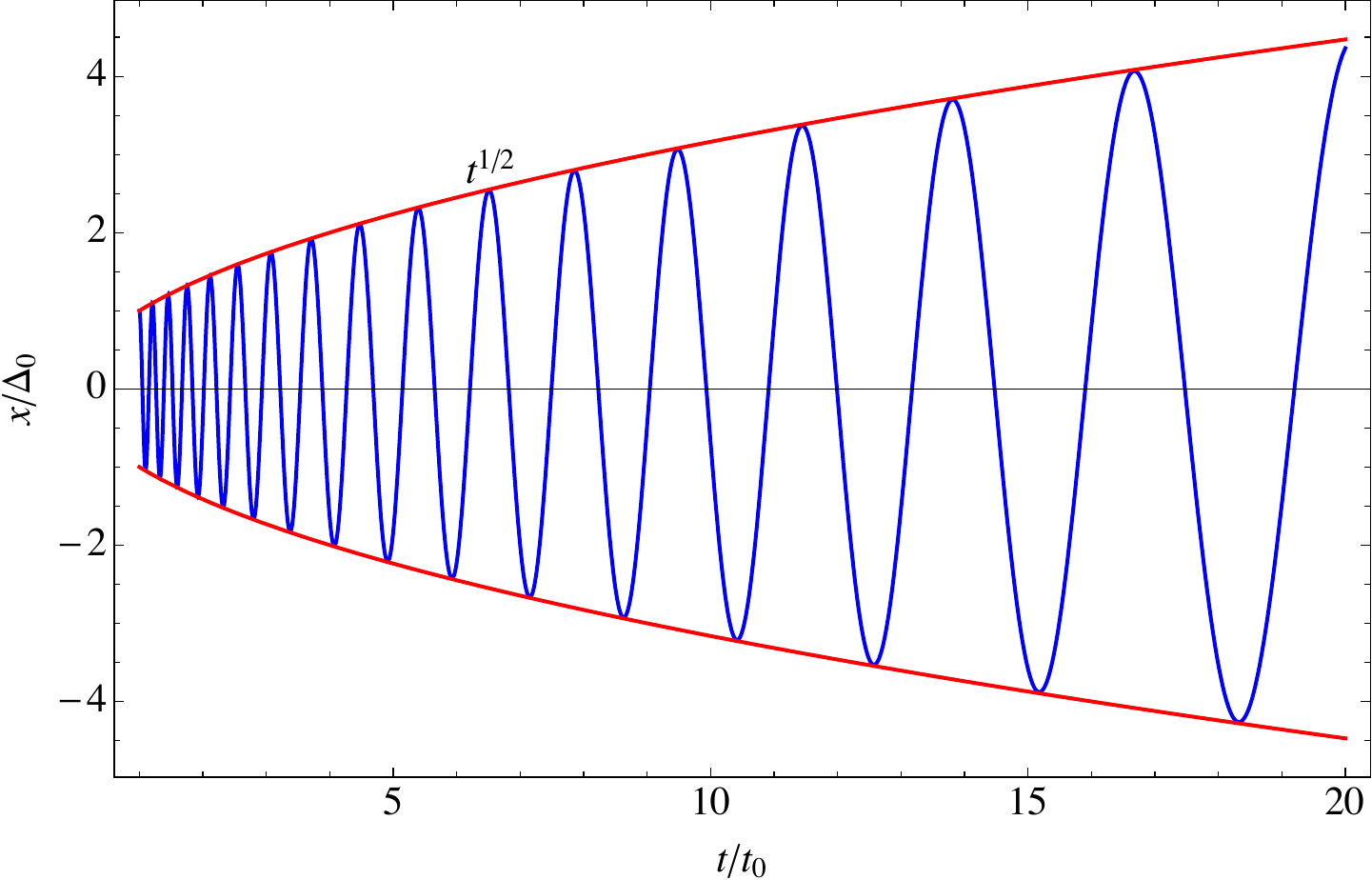}
\caption{Comparison of numerical solution for motion of particle inside the
sheet (blue line) with analytical prediction (\ref{eq:orb_orth}). Red line 
shows the dependence of amplitude of particle oscillations on time
$x_{\rm max} = x_0 t^{1/2}/t_0^{1/2}$.}
\label{fig:orb_orth}
\end{figure}

Expanding now r.h.s. into Maclaurin series (which is possible for $S \ll 1$) we obtain for the
leading term
\begin{equation}
\label{eqn.S.ort}
t^{2}S''+ 2\kappa t S' + \log (2) \Lambda^2S = 0.
\end{equation}
Here we suppose that $\Lambda \simeq 4\lambda\sigma_{\rm M} \gg 1$.
Equation (\ref{eqn.S.ort}) is linear and have the following solutions
\begin{equation}
\label{sol.S.ort}
S_{\pm} = S_0 \, t^{-(2\kappa-1)/2} 
\cos\left[\sqrt{\log (2)}\Lambda \log (t) + \varphi_{\pm}\right].
\end{equation}
As a result, it becomes clear that unless $\kappa\neq 1/2$ the expansion of the sheet 
doesn't follow the expansion of particle trajectory, for which we have
\begin{equation}
    x(t)\approx 
x_{0}\left(\frac{t}{t_0}\right)^{1/2}\cos\left[\alpha (x_0) \Lambda\log (t)
+\varphi_{0}\right]
\label{eq:orb_orth}
\end{equation}
for arbitrary oscillation amplitude $x_0$. The constant $\alpha(x_0)$ slowly varies
between $\sqrt{\log (2)}$ at $x_0 \ll \Delta$ and unity for \mbox{$x_0 \gg \Delta$.} 
An analytical solution (\ref{eq:orb_orth}) is compared with numerical solution in 
Figure~\ref{fig:orb_orth} demonstrating their good agreement. { It is also clear 
from (\ref{sol.S.ort}), that the growth rate proportional to square root of time is 
a solution for arbitrary expanding sheet.}

\subsection{Current sheet thickness and particle acceleration}
\label{sect.Est}

In previous section we have found self-consistent solution for internal 
structure of current sheet, however, in this section we will show, that 
such solution cannot be asymptotic one. Below we evaluate the current 
sheet thickness based on its global structure. We will explore MHD equations 
in order to find asymptotic behaviour of current sheet.

Starting from the Faraday's law
\begin{equation}
    \dfrac{B}{\Delta}=8\pi n_{\rm in} e \dfrac{v_z}{c}
\end{equation}
and combining it with the definition of multiplicity $\lambda$, we can express 
thickness of the current sheet through $v_{z}(t)/c$ and $n_{\rm out}/n_{\rm in}$ 
parameters:
\begin{equation}
\Delta=R_{L}\dfrac{t}{t_0}\dfrac{\Gamma}{4\lambda}\dfrac{c}{v_{z}}\dfrac{n_{\rm out}}{n_{\rm in}}.
\label{DD}
\end{equation}
Here $n_{\rm out}$ and $n_{\rm in}$ refer to particle number density outside a
nd inside the current sheet respectfully.

It is clear that asymptotically $v_{z}/c$ growths up to a constant value. For 
$v_{z}(\infty)/c<1$ we get non-relativistic case, which correspond to particle
motion discussed above. It is possible to determine value of 
$\delta p_{z}\simeq eB\Delta/c$:
 \begin{equation}
 \delta p_z = m_{\rm e}c \Gamma^2 \dfrac{c}{v_{z}}\dfrac{n_{\rm out}}{n_{\rm in}}.
 \label{eq:pz_final}
 \end{equation}
 This expression could be used for $\kappa \ge 1$ to get $\delta p_z(t)$, or for 
 $\kappa < 1$ to get $\delta p_z(\infty)$. For the latter case, the values of 
 $v_z$, $n_{\rm in}$ and $n_{\rm out}$ should be taken at $t = t_0$. 
 
 From equation (\ref{eq:pz_final}) one can see that if the sheet is relativistic 
 ($v_z \sim c$) and $n_{\rm in}\sim n_{\rm out}$, the particle can reach a Lorentz 
 factor $\gamma_{\rm e} \sim \Gamma^2 = \sigma_{\rm M}^{2/3}$. It is obvious that 
 such rapid acceleration breaks some of the assumption made in the beginning (e.g. 
 the Lorentz factor of the current sheet in the laboratory frame is only $\Gamma$).
 On the other hand, one may expect that in realistic sheet $n_{\rm in} \gg n_{\rm out}$, 
 so $\gamma_{\rm e} \ll \Gamma^2$. 
 
 { Further, as was shown earlier, the sheet with $\Delta \gg \mathcal{R}_{\rm L}$ 
 collapses. Under this condition the force balance perpendicular to the sheet plane 
 can be written as
 \begin{equation}
 \frac{B^2}{8\pi} = n_{\rm in} \mathcal{E}_{\rm k},
 \label{balance}
 \end{equation}
 where $\mathcal{E}_{\rm k}(t)$ is the  particle thermal kinetic energy inside the sheet. 
 Using now the definition (\ref{sigma}) for magnetization parameter $\sigma_{\rm M}$, 
 one can rewrite this relation in a simple form
 \begin{equation}
 \frac{\mathcal{E}_{\rm k}}{m_{\rm e} c^2} = 
 \frac{\sigma_{\rm M}}{\Gamma} \, \frac{n_{\rm out}}{n_{\rm in}}.
 \label{bbalance}
 \end{equation}
As we see, non-relativistic approximation inside the sheet is valid for high enough 
number density 
 \begin{equation}
  \frac{n_{\rm in}}{n_{\rm out}} > \sigma_{\rm M}^{2/3}.
 \label{nonrel}
 \end{equation}

As a result, comparing relations (\ref{DD}) and (\ref{bbalance}) one can conclude 
that linear increase of the sheet thickness $\Delta \propto t$ predicted by~\citet{coroniti_striped_wind_1990} and~\citet{michel_pulsar_wind_1994} and 
recently reproduced in numerical simulation~\citep{Q} can be realized for constant
particle energy $\mathcal{E}_{\rm k}$ and constant ratio $n_{\rm in}/n_{\rm out}$
only. The first condition $\mathcal{E}_{\rm k} \approx$ const is in agreement with 
time-independent acceleration energy along the sheet. On the other hand, the second
one $n_{\rm in}/n_{\rm out} \approx$ const can be realized only if the particle
inflow significantly increases the number of particles inside the sheet. Such as
inflow was also reproduced in numerical simulations (see e.g.~\citealt{2015MNRAS.448..606C}).
 
 Besides, if the plasma is hot, $\mathcal{E}_{\rm k} > m_{\rm e} c^2$, we obtain for 
 the Larmor radius
 \begin{equation}
 \langle\mathcal{R}_{\rm L} \rangle = \frac{c\langle p_\perp\rangle }{eB} =
 \frac{\mathcal{E}_{\rm k}}{eB} =
 R_{L}\dfrac{t}{t_0}\dfrac{\Gamma}{4\lambda}\dfrac{n_{\rm out}}{n_{\rm in}}.
 \end{equation}
As one can see, $\langle\mathcal{R}_{\rm L}\rangle = \Delta\cdot v_z/c$. Thus, for 
the condition  $\Delta \sim \mathcal{R}_{\rm L}$ to be fulfilled, relativistically 
hot current sheet requires $v_z \sim c$. 
 }
 
 { On the other hand, if the number density outside the sheet is much 
 smaller than inside, the number of particles inside the sheet cannot increase. This 
 will lead, to $n_{\rm in}\propto t^{-3}$, but $n_{\rm out}$ 
 will be proportional to $t^{-2}$. So the ratio between $n_{\rm out}$ and $n_{\rm in}$ 
 should be of order unity.}
 In this case particle acceleration, can be estimated as
\begin{equation}
    \delta p_z\simeq m_{\rm e}c \Gamma^2,
\end{equation}
or, in laboratory frame,
\begin{equation}
    \delta p_{z}^{\rm lab} \simeq m_{\rm e}c \Gamma.
\end{equation}
 Which is the same order as mean thermal momentum of particles inside current sheet.
 
 

\section{Conclusions and discussion}
\label{sect:conc}

{ This paper provides the formalism to describe essentially time-dependent 
evolution of the current sheet in the pulsar wind. As the first step in this direction, 
we carry out the calculations in the comoving reference frame and successfully determine 
intrinsic electromagnetic fields of the current sheet. In our opinion, this approach 
allows to describe more vividly the physical processes in a sheet.}


In the first part of this paper we investigate asymptotic structure of the wind from 
rotating oblique neutron star using force-free approximation. {  General asymptotic 
solution of the Grad-Shafranov equation for quasi-spherical pulsar wind up to the second 
order in small parameter \mbox{$\varepsilon = (\Omega r/c)^{-1}$} was  obtained. We have 
shown that the wind can have arbitrary latitude dependence of the energy flux. In particular, 
our solution describes the latitudinal structure of the radial magnetic field obtained 
numerically for oblique rotator. The form of current sheet in asymptotic does not depend on latitudinal structure of and matches the one in Bogovalov solution (\citealt{1999A&A...349.1017B}) } 

{ As the force-free approximation does not allow us to discuss the inner structure 
of the current sheet, we use MHD approximation in which the velocity of the pulsar wind 
is assumed to be less than that of light. Indeed, as it is well known (see, 
e.g.,~\citealt{bes98}), outside the fast magnetosonic surface the velocity of 
quasi-spherical MHD flow becomes almost constant. Using this property (and carrying 
out calculations in the comoving reference frame) we estimate the efficiency of the 
particle acceleration inside the sheet.}

{ The main conclusion of our consideration is that the intrinsic time-dependence 
of a sheet in the comoving reference frame (especially the increase of the sheet thickness 
$\Delta$) inevitably results in the appearance of the electric field which is larger than
magnetic one inside the sheet. It is this electric field that controls the electric current
of a sheet.}

{ Finally, after investigating the motion of individual particles in the time-dependent 
current sheet, we { evaluate} the width of the sheet and its time evolution. In particular, 
we considered both relativistic and non-relativistic temperatures inside the sheet.} 

{ In particular, it was shown, that while the individual particle orbit grows as 
$t^{1/2}$, the sheet as whole should growth linearly with time. This contradiction 
can be solved by using methods of kinetic equation. Since the particle inflow inside 
a sheet due to its expansion is considerable, the evolution of current sheet is 
determined by incoming particles and not by the evolution of individual particles 
inside the sheet.}



{ As for particle acceleration, we show that in relativistic case when number
density inside the sheet is similar to one outside the sheet, particle gains additional
$m_{\rm e}c\sigma^{1/3}$ momentum (in laboratory frame) due to expansion of a sheet. It is
important to notice that although the acceleration region $|{\boldsymbol E}| > |{\boldsymbol B}|$ is
narrow, the trajectories of particles are adjusted in such a way that they accelerate 
in MHD region $|{\boldsymbol E}| < |{\bf B}|$ every half period in one direction, cross 
$|{\bf E}| > |{\boldsymbol B}|$ zone, and then accelerate in the same direction again 
(see Figure~\ref{fig:trap}). }

Finally, it is important to highlight the difference between present article and~\citet{2001PASA...18..415K} in which authors use global equation in order to 
determine growth of current sheet thickness. In our work we do not consider 
reconnection, so our result should be interpreted as pre-reconnection sheet structure.
{ However, the electric field, which results from time-dependence of sheet 
thickness in case of linear growth, has the same structure as reconnection field  
and it is possible, that even in case of reconnection, our description of 
individual particle motion remains relevant.}

\section{Acknowledgments}

We thank Ya.N.~Istomin and A.Philippov for their interest and useful discussion 
{ and the anonymous referee for instructive comments which helped us to improve 
the manuscript}. This research was partially supported by the government of the 
Russian Federation (agreement No. 05.Y09.21.0018) and by Russian Foundation for 
Basic Research (Grant no. 15-02-03063).


{\small
\bibliographystyle{mn2e}
\bibliography{mybib}

\begin{thebibliography}{}

\bibitem[\protect\citeauthoryear{{Aharonian}, {Akhperjanian}, {Bazer-Bachi} \&
  {\etal}}{{Aharonian} et~al.}{2007}]{2007A&A...464..235A}
{Aharonian} F.,  {Akhperjanian} A.~G.,  {Bazer-Bachi} A.~R.,    {\etal} 2007,
  \aap, 464, 235

\bibitem[\protect\citeauthoryear{{Aharonian}, {Bogovalov} \&
  {Khangulyan}}{{Aharonian} et~al.}{2012}]{2012Natur.482..507A}
{Aharonian} F.~A.,  {Bogovalov} S.~V.,    {Khangulyan} D.,  2012, \nat, 482,
  507

\bibitem[\protect\citeauthoryear{{Arons}}{{Arons}}{1981}]{1981ApJ...248.1099A}
{Arons} J.,  1981, \apj, 248, 1099

\bibitem[\protect\citeauthoryear{{Arons}}{{Arons}}{2012}]{2012SSRv..173..341A}
{Arons} J.,  2012, \ssr, 173, 341

\bibitem[\protect\citeauthoryear{{Beskin}}{{Beskin}}{2010}]{2010mfca.book.....B}
{Beskin} V.~S.,  2010, {MHD Flows in Compact Astrophysical Objects}.
Springer

\bibitem[\protect\citeauthoryear{{Beskin}, {Gurevich} \& {Istomin}}{{Beskin}
  et~al.}{1993}]{BGI}
{Beskin} V.~S.,  {Gurevich} A.~V.,    {Istomin} Y.~N.,  1993, {Physics of the
  pulsar magnetosphere}.
{Cambridge University Press}

\bibitem[\protect\citeauthoryear{{Beskin}, {Kuznetsova} \& {Rafikov}}{{Beskin}
  et~al.}{1998}]{bes98}
{Beskin} V.~S.,  {Kuznetsova} I.~V.,    {Rafikov} R.~R.,  1998, \mnras, 299,
  341

\bibitem[\protect\citeauthoryear{{Beskin} \& {Rafikov}}{{Beskin} \&
  {Rafikov}}{2000}]{2000MNRAS.313..433B}
{Beskin} V.~S.,  {Rafikov} R.~R.,  2000, \mnras, 313, 433

\bibitem[\protect\citeauthoryear{{Beskin}, {Zakamska} \& {Sol}}{{Beskin}
  et~al.}{2004}]{beszak04}
{Beskin} V.~S.,  {Zakamska} N.~L.,    {Sol} H.,  2004, \mnras, 347, 587

\bibitem[\protect\citeauthoryear{{Blandford}}{{Blandford}}{1975}]{1975MNRAS.170..551B}
{Blandford} R.~D.,  1975, \mnras, 170, 551

\bibitem[\protect\citeauthoryear{{Bogovalov} \& {Tsinganos}}{{Bogovalov} \&
  {Tsinganos}}{1999}]{1999MNRAS.305..211B}
{Bogovalov} S.,  {Tsinganos} K.,  1999, \mnras, 305, 211

\bibitem[\protect\citeauthoryear{{Bogovalov}}{{Bogovalov}}{1999}]{1999A&A...349.1017B}
{Bogovalov} S.~V.,  1999, \aap, 349, 1017

\bibitem[\protect\citeauthoryear{{Bogovalov} \& {Khangoulyan}}{{Bogovalov} \&
  {Khangoulyan}}{2002}]{2002AstL...28..373B}
{Bogovalov} S.~V.,  {Khangoulyan} D.~V.,  2002, Astronomy Letters, 28, 373

\bibitem[\protect\citeauthoryear{{Brennan} \& {Gralla}}{{Brennan} \&
  {Gralla}}{2014}]{2014PhRvD..89j3013B}
{Brennan} T.~D.,  {Gralla} S.~E.,  2014, \prd, 89, 103013

\bibitem[\protect\citeauthoryear{{Bucciantini}, {Thompson}, {Arons}, {Quataert}
  \& {Del Zanna}}{{Bucciantini} et~al.}{2006}]{2006MNRAS.368.1717B}
{Bucciantini} N.,  {Thompson} T.~A.,  {Arons} J.,  {Quataert} E.,    {Del
  Zanna} L.,  2006, \mnras, 368, 1717

\bibitem[\protect\citeauthoryear{{Cerutti}, {Philippov}, {Parfrey} \&
  {Spitkovsky}}{{Cerutti} et~al.}{2015}]{2015MNRAS.448..606C}
{Cerutti} B.,  {Philippov} A.,  {Parfrey} K.,    {Spitkovsky} A.,  2015,
  \mnras, 448, 606

\bibitem[\protect\citeauthoryear{{Contopoulos}, {Kalapotharakos} \&
  {Kazanas}}{{Contopoulos} et~al.}{2014}]{2014ApJ...781...46C}
{Contopoulos} I.,  {Kalapotharakos} C.,    {Kazanas} D.,  2014, \apj, 781, 46

\bibitem[\protect\citeauthoryear{{Contopoulos}, {Kazanas} \&
  {Fendt}}{{Contopoulos} et~al.}{1999}]{ckf99}
{Contopoulos} I.,  {Kazanas} D.,    {Fendt} C.,  1999, \apj, 511, 351

\bibitem[\protect\citeauthoryear{Coroniti}{Coroniti}{1990}]{coroniti_striped_wind_1990}
Coroniti F.~V.,  1990, ApJ, 349, 538

\bibitem[\protect\citeauthoryear{{Gruzinov}}{{Gruzinov}}{2005}]{2005PhRvL..94b1101G}
{Gruzinov} A.,  2005, Physical Review Letters, 94, 021101

\bibitem[\protect\citeauthoryear{Harris}{Harris}{1962}]{harris1962}
Harris E.,  1962, Il Nuovo Cimento 1955-1965, 23, 115

\bibitem[\protect\citeauthoryear{{Ingraham}}{{Ingraham}}{1973}]{1973ApJ...186..625I}
{Ingraham} R.~L.,  1973, \apj, 186, 625

\bibitem[\protect\citeauthoryear{{Kalapotharakos}, {Contopoulos} \&
  {Kazanas}}{{Kalapotharakos} et~al.}{2012}]{2012MNRAS.420.2793K}
{Kalapotharakos} C.,  {Contopoulos} I.,    {Kazanas} D.,  2012, \mnras, 420,
  2793

\bibitem[\protect\citeauthoryear{{Komissarov}}{{Komissarov}}{2006}]{2006MNRAS.367...19K}
{Komissarov} S.~S.,  2006, \mnras, 367, 19

\bibitem[\protect\citeauthoryear{{Komissarov} \& {Lyubarsky}}{{Komissarov} \&
  {Lyubarsky}}{2003}]{2003MNRAS.344L..93K}
{Komissarov} S.~S.,  {Lyubarsky} Y.~E.,  2003, \mnras, 344, L93

\bibitem[\protect\citeauthoryear{{Levinson}}{{Levinson}}{2000}]{2000PhRvL..85..912L}
{Levinson} A.,  2000, Physical Review Letters, 85, 912

\bibitem[\protect\citeauthoryear{{Lyubarsky} \& {Kirk}}{{Lyubarsky} \&
  {Kirk}}{2001a}]{2001ApJ...547..437L}
{Lyubarsky} Y.,  {Kirk} J.~G.,  2001a, \apj, 547, 437

\bibitem[\protect\citeauthoryear{{Lyubarsky} \& {Kirk}}{{Lyubarsky} \&
  {Kirk}}{2001b}]{2001PASA...18..415K}
{Lyubarsky} Y.~E.,  {Kirk} J.~G.,  2001b, PASA, 18, 415

\bibitem[\protect\citeauthoryear{{Lyutikov}}{{Lyutikov}}{2011}]{2011PhRvD..83l4035L}
{Lyutikov} M.,  2011, \prd, 83, 124035

\bibitem[\protect\citeauthoryear{{Mestel}}{{Mestel}}{1973}]{1973Ap&SS..24..289M}
{Mestel} L.,  1973, \apss, 24, 289

\bibitem[\protect\citeauthoryear{{Michel}}{{Michel}}{1973}]{1973ApJ...180L.133M}
{Michel} F.~C.,  1973, \apjl, 180, L133

\bibitem[\protect\citeauthoryear{Michel}{Michel}{1994}]{michel_pulsar_wind_1994}
Michel F.~C.,  1994, ApJ, 431, 397

\bibitem[\protect\citeauthoryear{{Okamoto}}{{Okamoto}}{1974}]{1974MNRAS.167..457O}
{Okamoto} I.,  1974, \mnras, 167, 457

\bibitem[\protect\citeauthoryear{{P{\'e}tri}}{{P{\'e}tri}}{2013}]{2013MNRAS.434.2636P}
{P{\'e}tri} J.,  2013, \mnras, 434, 2636

\bibitem[\protect\citeauthoryear{{Philippov} \& {Spitkovsky}}{{Philippov} \&
  {Spitkovsky}}{2017}]{Q}
{Philippov} A.~A.,  {Spitkovsky} A.,  2017, \apj

\bibitem[\protect\citeauthoryear{{Prokofev}, {Arzamasskiy} \&
  {Beskin}}{{Prokofev} et~al.}{2015}]{beam2015}
{Prokofev} V.~V.,  {Arzamasskiy} L.~I.,    {Beskin} V.~S.,  2015, \mnras, 454,
  2146

\bibitem[\protect\citeauthoryear{{Ruderman} \& {Sutherland}}{{Ruderman} \&
  {Sutherland}}{1975}]{rs75}
{Ruderman} M.~A.,  {Sutherland} P.~G.,  1975, \apj, 196, 51

\bibitem[\protect\citeauthoryear{{Spitkovsky}}{{Spitkovsky}}{2006}]{2006ApJ...648L..51S}
{Spitkovsky} A.,  2006, \apjl, 648, L51

\bibitem[\protect\citeauthoryear{{Sturrock}}{{Sturrock}}{1971}]{sturrock71}
{Sturrock} P.~A.,  1971, \apj, 164, 529

\bibitem[\protect\citeauthoryear{{Tchekhovskoy}, {McKinney} \&
  {Narayan}}{{Tchekhovskoy} et~al.}{2009}]{2009ApJ...699.1789T}
{Tchekhovskoy} A.,  {McKinney} J.~C.,    {Narayan} R.,  2009, \apj, 699, 1789

\bibitem[\protect\citeauthoryear{{Tchekhovskoy}, {Philippov} \&
  {Spitkovsky}}{{Tchekhovskoy} et~al.}{2016}]{2016MNRAS.457.3384T}
{Tchekhovskoy} A.,  {Philippov} A.,    {Spitkovsky} A.,  2016, \mnras, 457,
  3384

\bibitem[\protect\citeauthoryear{{Tchekhovskoy}, {Spitkovsky} \&
  {Li}}{{Tchekhovskoy} et~al.}{2013}]{SashaMHD}
{Tchekhovskoy} A.,  {Spitkovsky} A.,    {Li} J.~G.,  2013, \mnras, 435, L1

\bibitem[\protect\citeauthoryear{{Timokhin}}{{Timokhin}}{2006}]{2006MNRAS.368.1055T}
{Timokhin} A.~N.,  2006, \mnras, 368, 1055

\bibitem[\protect\citeauthoryear{{Timokhin} \& {Arons}}{{Timokhin} \&
  {Arons}}{2013}]{2013MNRAS.429...20T}
{Timokhin} A.~N.,  {Arons} J.,  2013, \mnras, 429, 20

\bibitem[\protect\citeauthoryear{{Tomimatsu}}{{Tomimatsu}}{1994}]{1994PASJ...46..123T}
{Tomimatsu} A.,  1994, \pasj, 46, 123

\bibitem[\protect\citeauthoryear{{Zenitani} \& {Hoshino}}{{Zenitani} \&
  {Hoshino}}{2007}]{2007ApJ...670..702Z}
{Zenitani} S.,  {Hoshino} M.,  2007, \apj, 670, 702

\end{thebibliography}
}
\bsp
\appendix
\section{Fields in a comoving frame for align case}
\label{Appendix.align}

In this Appendix we investigate the particle motion inside and outside of
the time-dependent current sheet with magnetic field
\begin{align}
\label{eqn:B1}\\
B_y & = B_0 \frac{R_{\rm L}}{ct} \tanh\left[\frac{z}{\Delta(t)}\right]. 
\label{eqn:B2}
\end{align}

For the aligned case we consider the same functional form of the time dependence of the sheet thickness on time 
$\Delta(t)=\Delta_0(t/t_0)^{\kappa}$. Then, the expression for the electric field
\begin{eqnarray}
E_x & = & \dfrac{B_0 R_{\rm L} z}{c^2t^2} \tanh\left[\frac{z}{\Delta(t)}\right], 
\label{eqnB:exO2d}
\\
\label{Appendix:ez.al}
E_z & = & 0,
\end{eqnarray}
again does not satisfy Maxwell equations for arbitrary $\kappa$, and remains valid only for $\kappa = 1$. This corresponds to the linear growth of the sheet thickness due to the radial expansion with constant opening angle of a sheet.
 
Similarly to the orthogonal case there are two possibilities to make the fields satisfy Maxwell equations:

 (i) Changing $x$-component of electric field, without any change of $z$-component

 (ii) Changing $z$-component without any change of $x$-component.

In contrast to the orthogonal case we change $x$-component of the
electric field instead of adding $E_{z}$ one. This implies that Hamiltonian for aligned case 
will be independent of $x$:
\begin{equation}
\begin{split}
\mathcal{H}^{\rm al}
=c \left(m^2c^2+ P^{2}_{z} + p^{2}_{z}\right)^{1/2},
\end{split}
\end{equation}
where
\begin{equation}
P_{x} = p_{x} - \dfrac{eB_0R_L\Delta(t)}{c^2t}
\log\left[\cosh\left(\dfrac{z}{\Delta(t)}\right)\right].
\end{equation}

Finally, similarly to the orthogonal case, we can find the solution of Maxwell equations by choosing 
the integration constant such that electric field vanishes at infinity.
\begin{equation}
\label{Appendix:ex.al}
\begin{split}
E_{x} = \dfrac{B_0R_L\Delta(t)}{c^2t^2}
\left\{(1-\kappa)\log\left[2\cosh\left(\frac{z}{\Delta(t)}\right)\right]+\right.\\
\left.+\kappa\dfrac{z}{\Delta(t)}\tanh\left(\frac{z}{\Delta(t)}\right)\right\}.
\end{split}
\end{equation}
As a result, an estimation of the current sheet width based on the equality between 
Larmor radius and the sheet width is still valid for the aligned case. But here the drift leads  
to the rarefaction of sheet, since it is directed perpendicular to the sheet. On the other hand, since 
$\partial\mathcal{H}/\partial x =0$, we have $P_{x} =$ const. Hence, the particle can gain acceleration along the sheet only due to electromagnetic part of momentum
\begin{equation}
\begin{split}
\delta p_x=\dfrac{eB_0R_L}{c^2}\left\{\dfrac{\Delta_0}{t_0}
\log\left[\cosh\dfrac{z_0}{\Delta_0}\right]\right.
-\\-\left.
\dfrac{\Delta(t)}{t}
\log\left[\cosh\dfrac{z}{\Delta(t)}\right]\right\}
\end{split}
\end{equation}
 
\subsection{Particle motion}
\label{sect:al}

\begin{figure}
\centering
\includegraphics[scale=0.55]{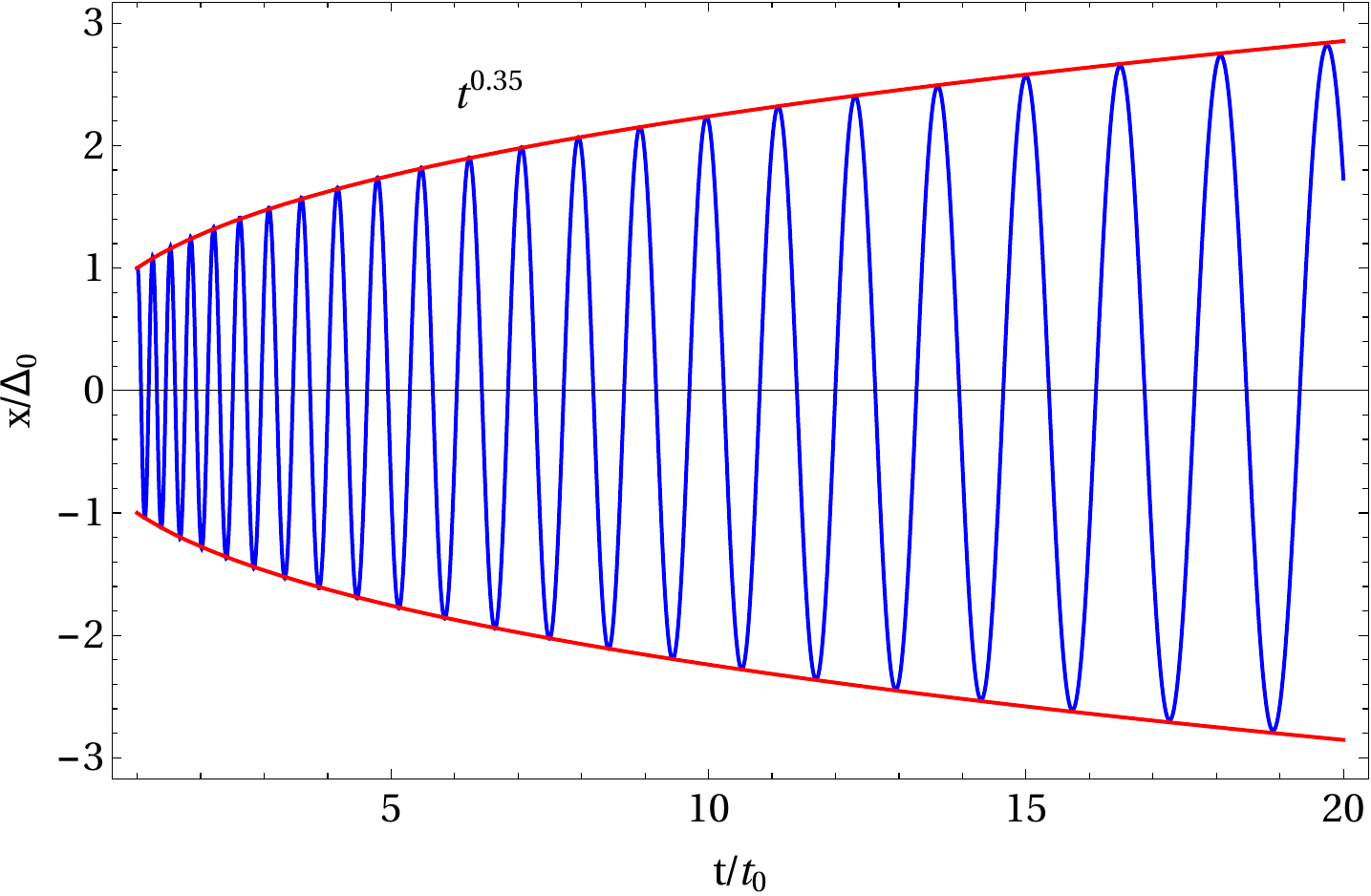}
\caption{Numerical solution to equation (\ref{App.case1}). The self consistency obtained for $\kappa=0.35$. }
\label{fig:App.sol}
\end{figure}

For aligned current sheet we use expressions (\ref{eqn:B2}), (\ref{Appendix:ez.al}), and
(\ref{Appendix:ex.al}) for electric and magnetic fields:
\begin{align}
& B_y   = B_0 \frac{t_0}{t} \tanh\frac{z}{\Delta(t)},\\
\begin{split}
 E_x = B_0\dfrac{R_{\rm L}\Delta(t)}{(ct)^2} 
\left\{(1-\kappa)\log\left[2\cosh\dfrac{z}{\Delta(t)}\right]+\right.\\
\left.+\kappa\dfrac{z}{\Delta(t)}\tanh\frac{x}{\Delta(t)}\right\},
\end{split}
\\
&E_z = 0.
\end{align}
Here again $\Delta(t)\propto t^{\kappa}$. After writing down equations of motion and some 
algebra we obtain the following equation for $z$ coordinate of the particle:
\begin{equation}
\ddot{z}=\dot{x}\dfrac{\Lambda}{t}\tanh\dfrac{z}{\Delta(t)}
\end{equation} 
 and equality for for $x$-component of velocity
 \begin{equation}
 \dot{x}=v_0+\Lambda\dfrac{\Delta_0}{t_0}\ln[2\cosh(x_0/\Delta_0)]-\Lambda\dfrac{\Delta}{t}\ln[2\cosh(x/\Delta)],
 \end{equation}
which corresponds to conservation of $x$-component of generalize momentum. 
After that we should consider three cases:
 \begin{enumerate}
 \item 
 $\kappa < 1$. In this case on large time-scales the first term dominates, 
so we can neglect the second one. The same takes place when $\kappa=1$ but $\dot{x}\sim {\rm const}\neq0$
\item
 $\kappa > 1$. In this case we neglect the first term.
\item
$\kappa = 1$, $\dot{x}\rightarrow 0$. This scenario is possible if $\Delta\propto t$. 
In this case $z(t)=z_0(t)+\delta z(t)$, where $z_0/\Delta =$ const.
Accordingly, $\delta z\ll z_0$ $z_0''(t)=0$.  
\end{enumerate}  
In the first case we have
  \begin{equation}
  \label{App.case1}
  \ddot{z}=\Lambda\dfrac{v_0+V}{t}\tanh(z/\Delta),
  \end{equation}
  where $V=\Lambda\dfrac{\Delta_0}{t_0}\ln[2\cosh(x/0/t_0)]$.
This equation cannot be solved analytically for arbitrary $\kappa$ even in $z/\Delta\ll 1$ limit. For $v+V_0<0$ the numerical solution of this equation has shown at (\ref{fig:App.sol}).

In second case we obtain following equation:
  \begin{equation}
\ddot{z} = -\Lambda^2\dfrac{\Delta}{t^2}\tanh(z/\Delta)\log[2\cosh(z/\Delta)].
\end{equation} 

This equation is the similar to (\ref{eq.x.full.ort}). As we already 
establish above, it describes particle orbit with $t^{1/2}$ radius growth rate.

Finally for $\kappa = 1$ we have
\begin{equation}
\ddot{\delta z}=-\Lambda^2\dfrac{\delta z}{t^2}\tanh^2(x_0/\Delta).
\end{equation}
The solution of this equation is 
\begin{equation}
\delta z=\sqrt{t}\cos\left[\Lambda
\tanh\left(\dfrac{x_0}{\Delta(t)}\right)\log t\right].
\end{equation}
  This solution describes radial motion of particle experiencing Larmor's motion in declining field.



\bsp
\label{lastpage}

\end{document}